\documentclass[aps,twocolumn]{revtex4} 	

\usepackage{graphicx}					
\usepackage{amssymb}
\usepackage{amsmath}
\usepackage{stmaryrd}
\usepackage{color}

\usepackage{epsfig}

\def\ov#1{\overline{#1}}

\def\wt#1{\widetilde{#1}}
\def\vb#1{\mbox{\boldmath$#1$}}
\def\pd#1#2{\frac{\partial #1}{\partial #2}}

\def\wh#1{\widehat{#1}}
\def\bdot{\,\vb{\cdot}\,}
\def\btimes{\,\vb{\times}\,}

\def\bhat{\wh{{\sf b}}}

\newcommand{\bc}{\begin{center}}
\newcommand{\ec}{\end{center}}
\newcommand{\bt}{\begin{tabbing}}
\newcommand{\et}{\end{tabbing}} 
\newcommand{\be}{\begin{eqnarray*}}
\newcommand{\ee}{\end{eqnarray*}}

\begin{document}

\title{Perturbative variational formulation of \\ the Vlasov-Maxwell equations}

\author{Alain J.~Brizard}

\affiliation{Department of Physics, Saint Michael's College, Colchester, VT 05439, USA}

\begin{abstract}
The perturbative variational formulation of the Vlasov-Maxwell equations is presented up to third order in the perturbation analysis. From the second and third-order Lagrangian densities, respectively, the first-order and second-order Vlasov-Maxwell equations are expressed in gauge-invariant and gauge-independent forms. Upon deriving the reduced second-order Vlasov-Maxwell Lagrangian for the linear nonadiabatic gyrokinetic Vlasov-Maxwell equations, the reduced Lagrangian densities for the linear drift-wave equation and the linear hybrid kinetic-magnetohydrodynamic (MHD) equations are derived, with their associated wave-action conservation laws obtained by Noether method. The exact wave-action conservation law for the linear hybrid kinetic-MHD equations is written explicitly. Lastly, a new form of the third-order Vlasov-Maxwell Lagrangian is derived in which ponderomotive effects play a crucial role.
\end{abstract}

\date{\today}

\maketitle

\section{Introduction}

The dynamical reduction of the Vlasov-Maxwell equations provides a systematic pathway toward the formal derivation of the nonlinear gyrokinetic Vlasov-Maxwell equations, which are used extensively in the investigation of the turbulent evolution of fusion magnetized plasmas \cite{Brizard_Hahm_2007,Garbet_2010,Krommes_2012}. The modern derivation of the gyrokinetic Vlasov-Maxwell equations \cite{Brizard_Hahm_2007} is based on a series of phase-space transformations generated by a canonical generating function $S$, which succeeds in decoupling the fast gyromotion from the intermediate bounce/transit motion along the field lines and the slow drift motion across the field lines. 

The purpose of the present paper is to explore the perturbative variational formulation of the exact and reduced Vlasov-Maxwell equations, from which exact conservation laws for the linearized and nonlinear Vlasov-Maxwell equations are derived by Noether method \cite{Noether}. In particular, we derive the exact wave-action conservation laws for the linear exact and reduced Vlasov-Maxwell equations without requiring the WKB approximation (as is assumed in the standard derivation \cite{Andrews_McIntyre_1978}). It is important to note, however, that these conservation laws are exact only within the limits in which they are derived, i.e., they will not be valid whenever higher-order effects \textcolor{red}{(or additional physics not included in the model)} must be taken into account.

\subsection{Geometric Lie-transform perturbation theory}

It was previously shown \cite{Brizard_2001} that perturbed Hamiltonian dynamics can be represented geometrically in terms of two Hamiltonian functions, with the generating function $S$ acting as the Hamiltonian for the perturbation evolution
\begin{equation} 
dz^{\alpha}/d\epsilon \equiv  \{ z^{\alpha},\;S\},
\label{eq:Ham_S}
\end{equation}
where perturbations are now treated as a continuous process, and the Hamiltonian $H$ acting as the generating function for infinitesimal canonical transformations described by the standard canonical Hamilton equations 
\begin{equation} 
dz^{\alpha}/d t \equiv \{ z^{\alpha},\; H\}. 
\label{eq:Ham_h}
\end{equation}
Both Hamiltonian functions $H$ and $S$ (which has units of action since $\epsilon$ is dimensionless) depend on the canonical phase-space coordinates ${\bf z} = ({\bf x},{\bf p})$, the time $t$, and the perturbation variable $\epsilon$ (with $\epsilon = 0$ representing an arbitrary reference state).

The condition that the two Hamiltonian operators $d/dt \equiv \partial/\partial t + \{\;,\;H\}$ and $d/d\epsilon \equiv \partial/\partial\epsilon + \{\;,\;S\}$ commute (i.e., the order of temporal and perturbative evolutions is immaterial) yields the relation
\begin{eqnarray}
0 & = & \left[\frac{d}{dt},\; \frac{d}{d\epsilon}\right]f \;=\; \frac{d}{dt}\left(\frac{df}{d\epsilon}\right) \;-\; \frac{d}{d\epsilon}\left(\frac{df}{dt}\right) \nonumber \\
 & = & \left\{ f,\; \left(\pd{S}{t} \;-\; \pd{H}{\epsilon} \;+\; \{S,\; H \} \right) \right\},
 \label{eq:commutator}
 \end{eqnarray}
where the function $f({\bf z},t,\epsilon)$ is arbitrary. Here, we used the definitions
\begin{eqnarray*}
\frac{d}{dt}\left(\frac{df}{d\epsilon}\right) & = & \frac{\partial^{2}f}{\partial t\partial\epsilon} + \left\{ \pd{f}{t}, S\right\} + \left\{ f, \pd{S}{t}\right\} \\
 &  &+\; \left\{ \pd{f}{\epsilon} + \{ f, S\}, H \right\}, \\
\frac{d}{d\epsilon}\left(\frac{df}{dt}\right) & = & \frac{\partial^{2}f}{\partial\epsilon\partial t} + \left\{ \pd{f}{\epsilon}, H\right\} + \left\{ f, \pd{H}{\epsilon}\right\} \\
 &  &+\; \left\{ \pd{f}{t} + \{ f, H\}, S \right\},
 \end{eqnarray*}
 and, after cancellations, we used the Jacobi property of the Poisson bracket: $\left\{ \{ f, S\},\frac{}{} H \right\} + \left\{ \{ H, f\},\frac{}{} S \right\} = -\,\left\{ \{ S, H\},\frac{}{} f \right\}$, to obtain Eq.~\eqref{eq:commutator}. Since this relation must hold for any function $f$, we obtain the constraint between the Hamiltonians $S$ and $H$:
\begin{equation}
\pd{S}{t} \;-\; \pd{H}{\epsilon} \;+\; \left\{S,\frac{}{} H \right\} \;\equiv\; 0,
\label{eq:Sh_constraint}
\end{equation}
which appears prominently in Lie-transform Hamiltonian perturbation theory \cite{Cary_1981,Littlejohn_1982}. 

For practical applications of the Hamiltonian constraint \eqref{eq:Sh_constraint} in Vlasov-Maxwell theory, we now consider the following perturbation power expansions
\begin{equation}
S \;\equiv\; \sum_{n = 1}^{\infty}n\,\epsilon^{n-1}\,S_{n} 
\label{eq:S_epsilon}
\end{equation}
and
\begin{equation}
H \;=\; \frac{1}{2m}\,\left|{\bf p} - \frac{e}{c}\,{\bf A}\right|^{2} \;+\; e\,\Phi \;\equiv\; \sum_{n = 0}^{\infty}\epsilon^{n}\,H_{n},
\label{eq:h_epsilon}
\end{equation} 
where the expansion \eqref{eq:S_epsilon} simply mirrors the expansion $\partial H/\partial\epsilon = \sum_{n=1}n\,\epsilon^{n-1}H_{n}$. In addition, the electromagnetic potentials and fields are expanded as
\begin{equation} 
(\Phi,{\bf A};\; {\bf E},{\bf B}) \;\equiv\; \sum_{n=0}^{\infty}\epsilon^{n}\,(\Phi_{n},{\bf A}_{n};\; {\bf E}_{n},{\bf B}_{n}), 
\label{eq:phiA_EB_n}
\end{equation}
where ${\bf E}_{n} \equiv -\nabla\Phi_{n} - c^{-1}\partial{\bf A}_{n}/\partial t$ and ${\bf B}_{n} \equiv \nabla\btimes{\bf A}_{n}$ are derived from the electromagnetic potentials
$(\Phi_{n},{\bf A}_{n})$. By substituting these expansions into Eq.~\eqref{eq:Sh_constraint}, we recover the first two Lie-transform perturbation equations \cite{Littlejohn_1982}
\begin{eqnarray}
\frac{d_{0}S_{1}}{dt} & = & H_{1} \;=\; e\,\left( \Phi_{1} \;-\; \frac{{\bf v}_{0}}{c}\bdot{\bf A}_{1} \right), 
\label{eq:S1_dot} \\
\frac{d_{0}S_{2}}{dt} & = & H_{2} - \frac{1}{2}\,\{ S_{1},\; H_{1}\} \nonumber \\
 & = & e\,\left( \Phi_{2} \;-\; \frac{{\bf v}_{0}}{c}\bdot{\bf A}_{2} \right) \;+\; \frac{e^{2}}{2mc^{2}}\;|{\bf A}_{1}|^{2} \nonumber \\
  &  &-\; \frac{1}{2}\,\{ S_{1},\; H_{1}\},
\label{eq:S2_dot}
\end{eqnarray}
where $d_{0}/dt \equiv \partial/\partial t + \{\;,\; H_{0}\}$ is the unperturbed Hamiltonian evolution operator, expressed in terms of the unperturbed Hamiltonian $H_{0} \equiv m|{\bf v}_{0}|^{2}/2 + e\,\Phi_{0}$, where 
${\bf v}_{0} \equiv [{\bf p} - (e/c){\bf A}_{0}]/m$ denotes the unperturbed particle velocity. Here, we note that the evolution of $S_{2}$ explicitly involves the second-order potentials $(\Phi_{2}, {\bf A}_{2})$ as well as the quadratic ponderomotive Hamiltonian $-\,\frac{1}{2}\{ S_{1},\; H_{1}\}$, which involves the solution of the first-order equation $d_{0}S_{1}/dt = H_{1}$. These ponderomotive effects will appear prominently in the third-order action functional to be derived in Secs.~\ref{sec:perturbed_action} and \ref{sec:cubic}.

Before proceeding with our perturbation analysis of the Vlasov-Maxwell equations, however, we need to specify under what conditions this analysis may be valid. The use of perturbation methods has an extensive history in plasma physics \cite{Frieman_Rutherford_1964,Davidson_1972,Cary_1981,Littlejohn_1982,Lichtenberg_Lieberman_1983} and each application requires a specific ordering (i.e., the identification of a small dimensionless parameter $\epsilon$) based on the space-time-scale separation of the reference and perturbed Vlasov-Maxwell states. It is, therefore, useful to consider the first-order perturbed fields $(S_{1},\Phi_{1},{\bf A}_{1})$ to represent small-amplitude linear waves that perturb the Vlasov-Maxwell reference state, which will be described in terms of a second-order variational formulation. Hence, the stability of the reference state can be investigated directly from the second-order variational principle. \textcolor{red}{We note that, in order to derive a meaningful perturbation theory, we must exclude parametric resonances \cite{Landau_Lifshitz_1976} at all perturbation orders, since these resonances can easily lead to a breakdown in the perturbation ordering.} Nonlinear wave-particle and wave-wave interactions (e.g., weak turbulence) will naturally enter at the second order (and beyond) in the perturbation analysis \cite{Davidson_1972}, which will require at least a third-order variational formulation. It is the ultimate goal of this work to present a perturbative hierarchy for the Vlasov-Maxwell equations. While it is readily understood that the exact linear wave-action conservation laws derived from the second-order variational formulation are not to be taken literally, the wave-action density for each wave involved in nonlinear wave-wave interactions (e.g., three-wave interactions) is used as a field variable \cite{Coppi_etal_1969} in order to express the so-called wave kinetic equation.

\subsection{Organization}

The remainder of the paper is organized as follows. In Sec.~\ref{sec:perturbed_action}, we construct a perturbative action functional for the Vlasov-Maxwell equations by imposing the Lie-transform constraint \eqref{eq:Sh_constraint}. The Lagrange multiplier used with this constraint is the Vlasov distribution function, which allows us to express the perturbation expansion of the Vlasov distribution in powers of the scalar fields $(S_{1}, S_{2}, ...)$. In Sec.~\ref{sec:quadratic}, the second-order action functional is derived from the perturbative Vlasov-Maxwell action functional. The second-order action functional is quadratic in either the first-order fields $(S_{1},\Phi_{1},{\bf A}_{1})$, in the gauge-invariant form, or the first-order fields $(\vb{\xi}_{1} \equiv \partial S_{1}/\partial{\bf p}, {\bf E}_{1}, {\bf B}_{1})$, in the gauge-independent form. In the 
gauge-independent form (which can also be derived from the Low-Lagrangian formulation \cite{Low_1958}), the first-order polarization and magnetization appear explicitly in the first-order Maxwell equations as well as in the energy-momentum and wave-action conservation laws (derived by Noether method). 

In Sec.~\ref{sec:reduced}, we review the applications of the quadratic Vlasov-Maxwell action functional that lead to the variational formulations of the linear drift-wave equation and the kinetic-magnetohydrodynamic (MHD) equations. In particular, we expand our previous work on the kinetic-MHD equations \cite{Brizard_1996} and derive the exact kinetic-MHD wave-action conservation law for the general case of a time-dependent nonuniform bulk plasma. 

In Sec.~\ref{sec:cubic}, we present the third-order Vlasov-Maxwell action functional, which is given in gauge-invariant and gauge-independent forms. The gauge-invariant third-order action functional is the sum of terms that are cubic in the first-order fields $(S_{1},\Phi_{1},{\bf A}_{1})$ as well as ponderomotive terms involving the second-order fields $(S_{2},\Phi_{2},{\bf A}_{2})$, which are traditionally absent from all previous third-order action functionals (see, for example, Ref.~\cite{Boyd_Turner_1972}). These ponderomotive terms, however, play an integral part in the Lie-transform formulation of perturbed Vlasov-Maxwell theory \cite{Cary_Kaufman_1981}. In future work, this third-order Vlasov-Maxwell action functional will be explored for applications in nonlinear reduced gyrokinetic theory. Lastly, general expressions for the perturbed polarization and magnetization associated with the perturbed particle phase-space dynamics \eqref{eq:Ham_S} are presented in Sec.~\ref{sec:pol_mag} and our work is summarized in Sec.~\ref{sec:summary}.

\section{\label{sec:perturbed_action}Perturbative action functional}

In the present Section, we introduce the perturbative variational formulation of the Vlasov-Maxwell equations. We start with the perturbed Vlasov action functional
\[ {\cal A}_{{\rm V}\epsilon} \equiv \int_{0}^{\epsilon}d\sigma\int f\left( \pd{S}{t} - \pd{H}{\sigma} + \left\{ S,\frac{}{} H \right\} \right) d^{6}z dt, \]
where the perturbation parameter $\sigma$ is integrated from the reference state ($\sigma = 0$) to the physically-perturbed state ($\sigma = \epsilon$) and the Lagrange multiplier $f({\bf z}; t, \sigma)$ will be interpreted below as the Vlasov distribution function [see Eq.~\eqref{eq:Vlasov_h}]; here, summation over particle species is implicitly assumed. 

\subsection{Perturbed Vlasov equation}

The variation of ${\cal A}_{{\rm V}\epsilon}$ with respect to $f$ yields the constraint \eqref{eq:Sh_constraint}, the variation of ${\cal A}_{{\rm V}\epsilon}$ with respect to $S$ yields the standard Vlasov equation
\begin{equation}
\frac{df}{dt} \;=\; \pd{f}{t} \;+\; \left\{f,\frac{}{} H \right\} \;=\; 0,
\label{eq:Vlasov_h}
\end{equation}
and the variation of ${\cal A}_{{\rm V}\epsilon}$ with respect to the Hamiltonian $H$ yields
\begin{equation}
\frac{df}{d\sigma} \;=\; \pd{f}{\sigma} \;+\; \left\{f,\frac{}{} S \right\} \;=\; 0,
\label{eq:Vlasov_S}
\end{equation}
which shows how the Vlasov perturbations 
\begin{equation}
f - f_{0} \;\equiv\; \sum_{n=1}^{\infty}\sigma^{n}\,f_{n}
\label{eq:f_sigma}
\end{equation}
are generated by $S$. More explicitly, the first two terms of the Vlasov perturbation hierarchy \eqref{eq:Vlasov_S} are
\begin{equation}
\left. \begin{array}{rcl}
f_{1} & = & \{ S_{1},\; f_{0} \} \\
 &  & \\
f_{2} & = & \{ S_{2},\; f_{0} \} \;+\; \frac{1}{2}\,\{ S_{1},\; f_{1} \}
\end{array} \right\}.
\label{eq:f_epsilon}
\end{equation}
We note that the first-order expression $f_{1} = \{ S_{1}, f_{0}\}$ is also used by Morrison and Pfirsch \cite{MorrisonPfirsch_1990} in applying the quadratic free-energy method on the stability of Vlasov equilibria.

\subsection{Perturbed Maxwell equations}

Next, we turn our attention to the perturbed Maxwell equations. For this purpose, we introduce perturbed Vlasov-Maxwell action functional
\begin{eqnarray}
{\cal A}_{\epsilon} & \equiv & \int_{0}^{\epsilon}d\sigma\left[\int d^{6}z dt\; f\left( \pd{S}{t} - \pd{H}{\sigma} + \left\{ S,\frac{}{} H \right\} \right) \right] \nonumber \\
 &  &+\; \int_{0}^{\epsilon}d\sigma\left[\int d^{3}r dt \left( \frac{\bf E}{4\pi}\vb{\cdot}\pd{\bf E}{\sigma} - \frac{\bf B}{4\pi}\vb{\cdot}\pd{\bf B}{\sigma} \right) \right],
\label{eq:action_epsilon}
\end{eqnarray}
which is now a functional of the electromagnetic potentials $(\Phi,{\bf A})$, through the Hamiltonian $H$, and the electromagnetic potential perturbation derivatives $(\partial\Phi/\partial\sigma,\partial{\bf A}/\partial
\sigma)$. We note that $(H,{\bf E},{\bf B})$ also depend on the reference potentials $(\Phi_{0},{\bf A}_{0})$, which are functionally independent from the perturbation fields $(\partial\Phi/\partial\sigma,\partial{\bf A}/\partial
\sigma)$. A slightly different version of the perturbative action functional \eqref{eq:action_epsilon} was presented by Larsson \cite{Larsson_1993}, where the Maxwell part is expressed solely in terms of unperturbed fields $({\bf E}_{0},{\bf B}_{0})$ and first-order perturbation fields $({\bf E}_{1},{\bf B}_{1})$. Hence, in Larsson's theory, the Vlasov-Maxwell fields $(f,{\bf E},{\bf B})$ are not treated equally since $(\partial{\bf E}/\partial\sigma,
\partial{\bf B}/\partial\sigma) \equiv ({\bf E}_{1}, {\bf B}_{1})$ are truncated at the lowest order while $\partial f/\partial\sigma$ is expanded to all orders.

In the Vlasov part of the action functional \eqref{eq:action_epsilon}, the perturbation derivative 
$\partial H/\partial\sigma$ of Eq.~\eqref{eq:h_epsilon} is 
\begin{equation}
\pd{H}{\sigma} \;=\; e\;\pd{\Phi}{\sigma} \;-\; e\;\pd{\bf A}{\sigma}\bdot\frac{\bf v}{c},
\label{eq:h_sigma}
\end{equation}
where ${\bf v} \equiv [{\bf p} - (e/c)\,{\bf A}]/m$ denotes the particle velocity, while the Maxwell part in Eq.~\eqref{eq:action_epsilon} can be written as
\begin{eqnarray*} 
 &  &\int \frac{d^{3}r\,dt}{4\pi} \left( {\bf E}\bdot\pd{\bf E}{\sigma} - {\bf B}\bdot\pd{\bf B}{\sigma} \right) \\
  & = & \int \frac{d^{3}r\,dt}{4\pi} \left[\pd{\Phi}{\sigma}\left( \nabla\bdot\frac{}{}{\bf E}\right) + \pd{\bf A}{\sigma}\bdot\left(\frac{1}{c}\,\pd{\bf E}{t} - \nabla\btimes{\bf B}\right)\right] 
  \end{eqnarray*}
after integration by parts is performed. By replacing Eq.~\eqref{eq:h_sigma} into Eq.~\eqref{eq:action_epsilon}, variations of ${\cal A}_{\epsilon}$ with respect to the perturbation fields
$(\partial\Phi/\partial\sigma, \partial{\bf A}/\partial\sigma)$ now yield the Maxwell equations
\begin{equation}
\nabla\bdot{\bf E} \;=\; 4\pi\,\int_{\bf p}\,e\,f 
\label{eq:div_E}
\end{equation}
and
\begin{equation}
\nabla\btimes{\bf B} \;-\; \frac{1}{c} \pd{\bf E}{t} \;=\; \frac{4\pi}{c}\int_{\bf p}\,e\,{\bf v}\,f, 
\label{eq:E_dot}
\end{equation}
where the momentum integral $\int_{\bf p} \equiv \sum \int d^{3}p$ includes a sum over particle species. The remaining source-free Maxwell equations
\begin{equation}
\left. \begin{array}{rcl}
\nabla\bdot{\bf B} & = &  0 \\
\partial{\bf B}/\partial t & = &  -\,c\,\nabla\btimes{\bf E}
\end{array} \right\}
\end{equation}
follow from the definitions of the electromagnetic fields in terms of the potentials. Note that these equations include contributions from the reference fields $(f_{0},{\bf E}_{0},{\bf B}_{0})$. 

The variations of ${\cal A}_{\epsilon}$ with respect to the reference potentials $(\Phi_{0}, {\bf A}_{0})$ yield the perturbed Maxwell equations 
\begin{eqnarray}
\nabla\bdot\pd{\bf E}{\sigma} & = & 4\pi\,\int_{\bf p} e\;\{ S,\; f\} \nonumber \\
 & \equiv & 4\pi\,\int_{\bf p} e\;\pd{f}{\sigma}, \label{eq:div_E_sigma} \\
\nabla\btimes\pd{\bf B}{\sigma} - \frac{1}{c}\frac{\partial^{2}{\bf E}}{\partial t\partial\sigma} & = & 4\pi\,\int_{\bf p} \frac{e}{c}\left({\bf v}\;\{ S,\; f\} - \frac{e\,f}{mc}\pd{\bf A}{\sigma}\right) \nonumber \\
 & \equiv & 4\pi\,\int_{\bf p} \frac{e}{c}\;\pd{}{\sigma}\left(\frac{d{\bf x}}{dt}\;f\right), \label{eq:curl_B_sigma}
\end{eqnarray}
where $d{\bf x}/dt = \{ {\bf x},\; H\} = {\bf v}$ and $\partial{\bf v}/\partial\sigma = -(e/mc)\partial{\bf A}/\partial\sigma$. We will return to these perturbed Maxwell equations in Sec.~\ref{sec:pol_mag}, where we will show that Eqs.~\eqref{eq:div_E_sigma}-\eqref{eq:curl_B_sigma} can be written as
\begin{eqnarray}
\nabla\bdot\pd{\bf E}{\sigma} & \equiv & -\;4\pi\,\nabla\bdot\mathbb{P}_{\sigma}, \label{eq:div_E_P} \\
\nabla\btimes\pd{\bf B}{\sigma} - \frac{1}{c}\frac{\partial^{2}{\bf E}}{\partial t\partial\sigma} & \equiv & \frac{4\pi}{c}\,\pd{\mathbb{P}_{\sigma}}{t} + 4\pi\;\nabla\btimes\mathbb{M}_{\sigma}, \label{eq:curl_B_PM}
\end{eqnarray}
where expressions for the polarization $\mathbb{P}_{\sigma}$ and the magnetization $\mathbb{M}_{\sigma}$ will be given in Sec.~\ref{sec:pol_mag}.

\subsection{Expansion of the action functional}

We now express the action functional \eqref{eq:action_epsilon} as a perturbation power series 
\begin{equation}
{\cal A}_{\epsilon} = \sum_{n = 1}^{\infty}\,\epsilon^{n}\;{\cal A}_{n} \equiv \sum_{n = 1}^{\infty}\,\epsilon^{n}\;\left( \sum_{k=0}^{n-1}\,
{\cal A}_{n-k}^{(k)}\right), 
\label{eq:An_def}
\end{equation}
where the $n$th-order action functional ${\cal A}_{n}$ describes the perturbed Vlasov-Maxwell dynamics, with the functional term ${\cal A}_{n-k}^{(k)}$ explicitly depending on $\vb{\Psi}_{n-k} \equiv (S_{n-k};
\Phi_{n-k},{\bf A}_{n-k})$. The contributions from the $n$th-order fields $\vb{\Psi}_{n}$, therefore, appear in the functional term
\begin{eqnarray}
{\cal A}_{n}^{(0)} & \equiv & \int d^{6}z\,dt f_{0} \left[ \frac{d_{0}S_{n}}{dt} \;-\; e \left( \Phi_{n} \;-\; \frac{{\bf v}_{0}}{c}\bdot{\bf A}_{n} \right) \right] \nonumber \\
 &  &+\; \int\frac{d^{3}r\,dt}{4\pi} \left( {\bf E}_{0}\bdot{\bf E}_{n} \;-\frac{}{} {\bf B}_{0}\bdot{\bf B}_{n}\right),
\label{eq:A_n_0}
\end{eqnarray}
where ${\bf v}_{0} = [{\bf p} - (e/c){\bf A}_{0}]/m$ denotes the particle's reference velocity. We now show that ${\cal A}_{n}^{(0)} \equiv 0$ at all orders $n \geq 1$ if the reference state $(f_{0},{\bf E}_{0},{\bf B}_{0})$ satisfies the reference Vlasov-Maxwell equations. First, if we integrate by parts the first term in Eq.~\eqref{eq:A_n_0}, we obtain
\[ \int S_{n}\,\left(d_{0}f_{0}/dt\right)\,d^{6}z\,dt \;\equiv\; 0, \]
which follows from the unperturbed (reference) Vlasov equation for $f_{0}$. Next, if we substitute ${\bf E}_{n} \equiv -\,\nabla\Phi_{n} - c^{-1}\partial{\bf A}_{n}/\partial t$ and ${\bf B}_{n} \equiv \nabla\btimes{\bf A}_{n}$ into the second term in Eq.~\eqref{eq:A_n_0} and integrate by parts, we obtain 
\[ \int \Phi_{n} \left(\nabla\bdot{\bf E}_{0} - 4\pi\,\int_{\bf p}\,e f_{0} \right)\,d^{3}r\,dt \;\equiv\; 0 \]
and
\[ \int {\bf A}_{n}\bdot\left(\frac{1}{c}\pd{{\bf E}_{0}}{t} - \nabla\btimes{\bf B}_{0} + \frac{4\pi}{c}\int_{\bf p}e{\bf v}_{0}f_{0} \right) d^{3}rdt \equiv 0, \]
which follow from the unperturbed (reference) Maxwell equations for ${\bf E}_{0}$ and ${\bf B}_{0}$. Hence, the functional term \eqref{eq:A_n_0} vanishes identically and the 
$n$th-order action functional 
\begin{equation}
{\cal A}_{n} \;=\; \sum_{k=1}^{n-1}\,{\cal A}_{n-k}^{(k)} 
\label{eq:An_sum}
\end{equation}
depends explicitly on the perturbation fields $(\vb{\Psi}_{n-1},...,\vb{\Psi}_{2},\vb{\Psi}_{1})$, with ${\cal A}_{1} \equiv 0$ appearing as a special case. The $n$th-order action functional ${\cal A}_{n}$, therefore, describes the $(n-1)$th-order perturbed Vlasov-Maxwell dynamics (i.e., ${\cal A}_{2}$ describes linear Vlasov-Maxwell dynamics while ${\cal A}_{3}$ can be used to describe second-order ponderomotive-driven Vlasov-Maxwell equations). 

\section{\label{sec:quadratic}Second-order Lagrangian Density}

The simplest perturbative action functional in Eq.~\eqref{eq:An_def} therefore appears at the second order, where the (quadratic) action functional ${\cal A}_{2} \equiv \int {\cal L}_{2}\,d^{3}r\,dt$ describes the linear (first-order) perturbed Vlasov-Maxwell dynamics. Here, the quadratic Lagrangian density is defined as
\begin{eqnarray}
{\cal L}_{2} & \equiv & \int_{\bf p} \;\left[ f_{1} \left( \frac{1}{2}\,\frac{d_{0}S_{1}}{dt} \;-\; H_{1}\right) \;-\;
\frac{e^{2}\,f_{0}}{2\,mc^{2}}\;|{\bf A}_{1}|^{2} \right] \nonumber \\
 &  &+\; \frac{1}{8\pi} \left( |{\bf E}_{1}|^{2} \;-\frac{}{} |{\bf B}_{1}|^{2} \right),
\label{eq:action_quad}
\end{eqnarray}
which depends on the perturbed Vlasov distribution $f_{1} = \{ S_{1},\; f_{0}\}$ and the perturbed electromagnetic fields $({\bf E}_{1},{\bf B}_{1})$. The Eulerian variation of the Lagrangian density \eqref{eq:action_quad} is expressed as
\begin{widetext}
\begin{eqnarray}
\delta{\cal L}_{2} & = & \int_{\bf p} \;\left[ \{ \delta S_{1},\; f_{0}\} \left( \frac{1}{2}\,\frac{d_{0}S_{1}}{dt} \;-\; H_{1}\right) \;+\; f_{1} \left( \frac{1}{2}\,\frac{d_{0}\delta S_{1}}{dt} \;-\; \delta H_{1}\right) \;-\; \frac{e^{2}\,f_{0}}{mc^{2}}\;\delta{\bf A}_{1}\bdot{\bf A}_{1} \right] \nonumber \\
 &  &-\; \frac{1}{4\pi} \left[ {\bf E}_{1}\bdot\left(\nabla\delta\Phi_{1} \;+\; \frac{1}{c}\pd{\delta{\bf A}_{1}}{t}\right) \;+\; {\bf B}_{1}\bdot\nabla\btimes\delta{\bf A}_{1} \right] \nonumber \\
  & \equiv & \pd{\delta{\cal J}_{2}}{t} \;+\; \nabla\bdot\delta\vb{\Gamma}_{2} \;-\; \int_{\bf p} \delta S_{1} \left\{ \left(\frac{d_{0}S_{1}}{dt} - H_{1}\right),\; f_{0} \right\} \;+\; \frac{\delta\Phi_{1}}{4\pi} \left( \nabla\bdot{\bf E}_{1} \;-\; 4\pi\int_{\bf p} e\,f_{1} \right) \nonumber \\
   &  &+\; \frac{\delta{\bf A}_{1}}{4\pi}\bdot\left[ \frac{1}{c}\pd{{\bf E}_{1}}{t} - \nabla\btimes{\bf B}_{1} + 4\pi\int_{\bf p} \frac{e}{c}\,\left( f_{1}{\bf v}_{0} - \frac{e{\bf A}_{1}}{mc}f_{0} \right) \right],
 \label{eq:delta_L2}
\end{eqnarray}
\end{widetext}
where the second expression is obtained after rearranging terms in order to isolate the variations $(\delta S_{1}, \delta\Phi_{1},\delta{\bf A}_{1})$. We note that the space-time divergence terms (to be defined below) do not contribute in the quadratic variational principle \cite{Brizard_1994a} $\int \delta{\cal L}_{2}\,d^{3}r\,dt = 0$.

\subsection{First-order Vlasov-Maxwell equations}

Variation of the quadratic Lagrangian density \eqref{eq:delta_L2} with respect to $S_{1}$ yields the first-order Vlasov equation
\begin{equation}
0 = \frac{d_{0}f_{1}}{dt} \;+\; \{ f_{0},\; H_{1} \} = \left\{ \left(\frac{d_{0}S_{1}}{dt} - H_{1}\right),\; f_{0} \right\},
\label{eq:S1_dot_f0}
\end{equation}
which becomes
\begin{equation}
\frac{d_{0}S_{1}}{dt} \;=\; H_{1} \;=\; e\,\left( \Phi_{1} \;-\; {\bf A}_{1}\bdot\frac{{\bf v}_{0}}{c}\right),
\label{eq:S1_eq}
\end{equation}
when an arbitrary reference Vlasov distribution $f_{0}$ is considered (which satisfies $d_{0}f_{0}/dt = 0$). Variations of the quadratic Lagrangian density \eqref{eq:delta_L2} with respect to $(\Phi_{1},{\bf A}_{1})$ yield the first-order Maxwell equations
\begin{eqnarray}
\nabla\bdot{\bf E}_{1} & = & 4\pi\,\int_{\bf p}e\;f_{1}, 
\label{eq:div_E1} \\
\nabla\btimes{\bf B}_{1} - \frac{1}{c}\pd{{\bf E}_{1}}{t} & = & \frac{4\pi}{c}\int_{\bf p} e\left( f_{1}{\bf v}_{0} - \frac{e{\bf A}_{1}}{mc}f_{0} \right),
\label{eq:curl_B1}
\end{eqnarray}
with $f_{1} = \{ S_{1},\; f_{0}\}$. Equations \eqref{eq:S1_eq}-\eqref{eq:curl_B1} describe the standard linearized Vlasov-Maxwell equations, from which linear waves and instabilities in a general Vlasov-Maxwell equilibrium state can be analysed. 

\subsection{Quadratic conservation laws}

The conservation laws of energy-momentum, angular-momentum, and wave-action associated with the linear Vlasov-Maxwell equations can be derived by Noether method \cite{Noether} from  $\delta{\cal L}_{2}$ as follows. We note that, when the unperturbed Vlasov-Maxwell fields $(f_{0}; {\bf E}_{0},{\bf B}_{0})$ are time-dependent and spatially nonuniform \cite{BCK_1993}, only the quadratic wave action is conserved exactly, while the energy and momentum associated with the perturbation fields $(S_{1}, {\bf E}_{1},{\bf B}_{1})$ are no longer conserved, since energy-momentum is exchanged with the reference Vlasov-Maxwell plasma. 

To demonstrate the power of the Noether method, we introduce the quadratic Noether equation obtained from Eq.~\eqref{eq:delta_L2}:
\begin{equation}
\delta{\cal L}_{2} \;=\; \pd{\delta{\cal J}_{2}}{t} \;+\; \nabla\bdot\delta\vb{\Gamma}_{2},
\label{eq:L2_Noether}
\end{equation}
which is left in Eq.~\eqref{eq:delta_L2} after Eqs.~\eqref{eq:S1_eq}-\eqref{eq:curl_B1} are derived from the variational principle. Here, the Noether fields
\begin{eqnarray}
\delta{\cal J}_{2} & = & \frac{1}{2}\, \int_{\bf p} f_{1}\,\delta S_{1} \;-\; \frac{{\bf E}_{1}}{4\pi\,c}\bdot\delta{\bf A}_{1}, \label{eq:J_2} \\
\delta\vb{\Gamma}_{2} & = &  \frac{1}{2}\, \int_{\bf p}\delta S_{1} \left( f_{1}\;\pd{H_{0}}{\bf p} \;-\; H_{1}\;\pd{f_{0}}{\bf p}\right)  \nonumber \\
 &  &-\; \frac{1}{4\pi} \left( \delta\Phi_{1}\,{\bf E}_{1} \;+\frac{}{} \delta{\bf A}_{1}\btimes{\bf B}_{1} \right)
\label{eq:Gamma_2}
\end{eqnarray}
are expressed in terms of the field variations $\delta\vb{\Psi}_{1} = (\delta S_{1}, \delta\Phi_{1}, \delta{\bf A}_{1})$. The Noether method involves relating symmetries of the Lagrangian density 
${\cal L}_{2}$ with exact conservation laws of the linear Vlasov-Maxwell equations, which are obtained by expressing the field variations $\delta\vb{\Psi}_{1}$ in terms of space-time translations or rotations.

\subsubsection{Quadratic energy conservation law}

As an application of the Noether method, we consider the energy conservation law associated with the symmetry of the Lagrangian density ${\cal L}_{2}$ under infinitesimal time translation $t \rightarrow t + \delta t$. First, an infinitesimal time translation induces the Eulerian variations $\delta\vb{\Psi}_{1} = -\,\delta t\,\partial\vb{\Psi}_{1}/\partial t$, with $\delta{\bf A}_{1} \equiv c\,\delta t\,({\bf E}_{1} + \nabla\Phi_{1})$, and 
$\delta{\cal L}_{2} \equiv -\delta t\,(\partial/\partial t - \partial_{0}/\partial t){\cal L}_{2}$, where $\partial_{0}{\cal L}_{2}/\partial t$ represents the explicit time dependence associated with the unperturbed Vlasov-Maxwell fields $(f_{0}; {\bf E}_{0},{\bf B}_{0})$. Next, by inserting these variations into 
the Noether fields \eqref{eq:J_2}-\eqref{eq:Gamma_2}, we obtain
\begin{equation}
\pd{{\cal E}_{2}}{t} \;+\; \nabla\bdot{\bf S}_{2} \;=\; -\;\frac{\partial_{0}{\cal L}_{2}}{\partial t}.
\label{eq:energy_2}
\end{equation}
where the quadratic energy density is
\begin{eqnarray}
{\cal E}_{2} & = & \frac{1}{8\pi} \left( |{\bf E}_{1}|^{2} \;+\frac{}{} |{\bf B}_{1}|^{2} \right) \;+\;  \left( \int_{\bf p}\frac{e^{2}\,f_{0}}{2mc^{2}}\right) |{\bf A}_{1}|^{2} \nonumber \\
 &  &-\;  \int_{\bf p}\left[ f_{1}\left( \frac{1}{2}\{ S_{1}, H_{0}\} + \frac{e}{c}{\bf A}_{1}\bdot{\bf v}_{0}\right) \right],
 \label{eq:E2_def}
\end{eqnarray}
and the quadratic energy-density flux is
\begin{eqnarray}
{\bf S}_{2} & = & \frac{c{\bf E}_{1}}{4\pi}\btimes{\bf B}_{1} \;-\; \Phi_{1} \int_{\bf p} \left( e\,{\bf v}_{0}\,f_{1} \;-\; \frac{e^{2}\,f_{0}}{mc}\;{\bf A}_{1} \right) \nonumber \\
 &  &+\; \frac{1}{2}\int_{\bf p} \pd{S_{1}}{t} \left( f_{1}\;\pd{H_{0}}{\bf p} - H_{1}\;\pd{f_{0}}{\bf p}\right).
 \end{eqnarray}
Hence, according to the Noether Theorem, the quadratic energy \eqref{eq:E2_def} is conserved if the reference Vlasov-Maxwell fields are time-independent (i.e., $\partial_{0}{\cal L}_{2}/\partial t \equiv 0$). We note that when the quadratic energy energy density \eqref{eq:E2_def} is integrated over space, we recover the quadratic free energy ${\cal F}_{2} \equiv \int {\cal E}_{2}\,
d^{3}x$ derived by Morrison and Pfirsch \cite{MorrisonPfirsch_1990}.

\subsubsection{Quadratic wave-action conservation law}

While the quadratic energy ${\cal E}_{2}$ is no longer conserved when the reference Vlasov-Maxwell fields $(f_{0}; {\bf E}_{0},{\bf B}_{0})$ are time-dependent, however, it is possible to construct an exact quadratic wave-action conservation law \cite{BCK_1993} $\partial\ov{\cal J}_{2}/\partial t + \nabla\bdot\ov{\vb{\Gamma}}_{2} = 0$. First, we consider complex-valued wave-fields \cite{BCK_1993} with $\vb{\Psi}_{1}^{*} = (S_{1}^{*},\Phi_{1}^{*},{\bf A}_{1}^{*}) \neq \vb{\Psi}_{1}$, and construct real-valued (eikonal-averaged) expressions for the Noether densities \eqref{eq:J_2}-\eqref{eq:Gamma_2}. Next, we introduce the eikonal-phase-like variations $\delta\vb{\Psi}_{1} = i\,\delta\theta\,\vb{\Psi}_{1}$ and $\delta\vb{\Psi}_{1}^{*} = -\,i\,\delta\theta\,\vb{\Psi}_{1}^{*}$, which yield $\delta{\cal J}_{2} = -\,\delta\theta\,\ov{\cal J}_{2}$ and $\delta\vb{\Gamma}_{2} = -\,\delta\theta\,\ov{\vb{\Gamma}}_{2}$, where the quadratic wave-action density $\ov{\cal J}_{2}$ and wave-action-density flux $\ov{\vb{\Gamma}}_{2}$ are defined as
\begin{eqnarray}
\ov{\cal J}_{2} & \equiv & {\rm Im}\left[\frac{{\bf A}_{1}^{*}\bdot{\bf E}_{1}}{4\pi\,c} \;+\; \frac{1}{2} \int_{\bf p}\left\{ S_{1}^{*},\frac{}{} f_{0}
\right\}\; S_{1}\right], \label{eq:J2_action} \\
\ov{\vb{\Gamma}}_{2} & \equiv & {\rm Im}\left[\frac{1}{4\pi} \left(\Phi_{1}^{*}\,{\bf E}_{1} \;+\frac{}{} {\bf A}_{1}^{*}\btimes{\bf B}_{1} \right) \right] \label{eq:Gamma2_action} \\
 &  &+ \frac{1}{2} {\rm Im}\left[\int_{\bf p} \left(\left\{ S_{1}^{*},\frac{}{} f_{0} \right\}\pd{H_{0}}{\bf p} - H_{1}^{*}\pd{f_{0}}{\bf p}\right) S_{1}\right].
\nonumber
\end{eqnarray}
Wave-action conservation laws play a crucial role, for example, in the linear mode conversion involving two coupled linear waves in a nonuniform background plasma \cite{Ray_2014}.

\subsection{Gauge-independent formulation}

We note that the quadratic Lagrangian density \eqref{eq:action_quad} is not gauge independent since the electromagnetic potentials $(\Phi_{1},{\bf A}_{1})$ appear explicitly in the first-order Hamiltonian \eqref{eq:S1_dot}. However, under the gauge transformation generated by an arbitrary gauge field $\chi_{1}({\bf x},t)$:
\begin{equation} 
(\Phi_{1},{\bf A}_{1},S_{1}) \rightarrow \left(\Phi_{1} - \frac{1}{c}\pd{\chi_{1}}{t}, {\bf A}_{1} + \nabla \chi_{1}, S_{1} - \frac{e}{c}\chi_{1}\right),
\label{eq:gauge_1}
\end{equation}
with the associated gauge transformations
\begin{equation}
\left( f_{1},\; H_{1} \right) \;\rightarrow\; \left( f_{1} - \frac{e}{c}\,\{\chi_{1},\; f_{0}\},\; H_{1} - \frac{e}{c}\,\frac{d_{0}\chi_{1}}{dt} \right), 
\label{eq:gauge_fh1}
\end{equation}
we can easily verify that Eq.~\eqref{eq:S1_eq} is gauge invariant, while the quadratic Lagrangian density \eqref{eq:action_quad} becomes ${\cal L}_{2} = {\cal L}_{2}^{\prime} + \partial\Lambda_{2}/\partial t + \nabla\bdot\vb{\Lambda}_{2}$, where $(\Lambda_{2}, \vb{\Lambda}_{2})$ are momentum integrals involving $(\chi_{1}, S_{1})$. Since the quadratic variational principle $\delta{\cal A}_{2} = 0$ is based on the action functional ${\cal A}_{2} = \int{\cal L}_{2}\,d^{3}x\,dt$, where the Lagrangian density ${\cal L}_{2}$ is integrated over space and time, then the action functional is invariant under the gauge transformations
\eqref{eq:gauge_1}-\eqref{eq:gauge_fh1}, since $\partial\Lambda_{2}/\partial t + \nabla\bdot\vb{\Lambda}_{2}$ is an exact space-time derivative. Similarly, the quadratic energy conservation law \eqref{eq:energy_2} is gauge invariant because, under a gauge transformation, the energy density ${\cal E}_{2}$ and energy-density flux ${\bf S}_{2}$ transform as ${\cal E}_{2} = {\cal E}_{2}^{\prime} + \nabla\bdot{\bf Q}_{2}$ and ${\bf S}_{2} = {\bf S}_{2}^{\prime} - \partial{\bf Q}_{2}/\partial t$, which leaves the quadratic energy conservation law \eqref{eq:energy_2} invariant.

We can eliminate all gauge dependence in what follows by introducing the gauge-independent first-order phase-space displacement 
\begin{equation}
\vb{\eta}_{1} \;\equiv\; \{{\bf x},\; S_{1} \} \;=\; \partial S_{1}/\partial{\bf p}, 
\label{eq:xi1_def}
\end{equation}
from which we define the gauge-invariant first-order velocity
\begin{equation}
{\bf u}_{1} \equiv \frac{d_{0}\vb{\eta}_{1}}{dt} - \vb{\eta}_{1}\bdot\nabla{\bf v}_{0} \;=\; -\; \frac{1}{m} \left( \nabla S_{1} \;+\; \frac{e}{c}\;{\bf A}_{1}\right),
\label{eq:xi1_dot}
\end{equation}
obtained from Eq.~\eqref{eq:S1_eq}, which satisfies the gauge-independent equation of motion \cite{Low_1958}
\begin{equation}
\frac{d_{0}^{2}\vb{\eta}_{1}}{dt^{2}} = \frac{e}{m}\,\left( {\bf E}_{1} + \frac{{\bf v}_{0}}{c}\btimes{\bf B}_{1}\right) + \frac{d_{0}\vb{\eta}_{1}}{dt}\btimes\frac{e{\bf B}_{0}}{mc},
\label{eq:xi1_ddot}
\end{equation}
where we have assumed uniform Maxwell fields $({\bf E}_{0},{\bf B}_{0})$ for simplicity. We note, here, that the first-order displacement \eqref{eq:xi1_def} is still a function on the full particle phase space. 

The first-order Maxwell equations \eqref{eq:div_E1}-\eqref{eq:curl_B1}, on the other hand, become
\begin{eqnarray}
\nabla\bdot{\bf E}_{1} & = & 4\pi\,\int_{\bf p}e\;f_{1} \equiv -\;4\pi\,\nabla\bdot\mathbb{P}_{1}, 
\label{eq:div_EP1} \\
\nabla\btimes{\bf B}_{1} - \frac{1}{c}\pd{{\bf E}_{1}}{t} & = & \frac{4\pi}{c}\int_{\bf p} e\left( f_{1}\,{\bf v}_{0} - \frac{e{\bf A}_{1}}{mc}\;f_{0} \right) \nonumber \\
 & \equiv & \frac{4\pi}{c}\;\pd{\mathbb{P}_{1}}{t} \;+\; 4\pi\;\nabla\btimes
\mathbb{M}_{1},
 \label{eq:curl_BM1}
\end{eqnarray}
where the first-order polarization and magnetization
\begin{equation}
\left( \mathbb{P}_{1},\; \mathbb{M}_{1}\right) \equiv \int_{\bf p} e\,f_{0} \left( \vb{\eta}_{1},\; \vb{\eta}_{1}\btimes\frac{{\bf v}_{0}}{c}\right)
\label{eq:PM_1}
\end{equation}
are defined in terms of moments of the first-order displacement $\vb{\eta}_{1}$, and the first-order magnetization is solely due to the moving electric-dipole contribution. Using the macroscopic fields ${\bf D}_{1} \equiv {\bf E}_{1} + 4\pi\,\mathbb{P}_{1}$ and ${\bf H}_{1} \equiv {\bf B}_{1} - 4\pi\,\mathbb{M}_{1}$, the first-order Maxwell equations \eqref{eq:div_EP1}-\eqref{eq:curl_BM1} become
\begin{equation}
\left. \begin{array}{rcl}
\nabla\bdot{\bf D}_{1} & = &  0 \\
c\,\nabla\btimes{\bf H}_{1} - \partial{\bf D}_{1}/\partial t & = &  0
\end{array} \right\}.
\label{eq:Maxwell_DH}
\end{equation}
Hence, in general first-order Vlasov-Maxwell theory, the perturbed first-order charge and current densities are entirely expressed in terms of perturbed first-order polarization charge and polarization/magnetization current densities, respectively. See the case of the oscillation-center Vlasov-Maxwell equations \cite{Brizard_2009} as an explicit example.

The first-order Vlasov-Maxwell equations \eqref{eq:xi1_ddot}-\eqref{eq:curl_BM1} can be obtained from the gauge-independent Lagrangian density
\begin{eqnarray}
{\cal L}^{\prime}_{2} & = & \frac{1}{2} \int_{\bf p} f_{0} \left[ m\,\left|\frac{d_{0}\vb{\eta}_{1}}{dt}\right|^{2} \;+\; \frac{e}{c}\left(\vb{\eta}_{1}\btimes\frac{d_{0}\vb{\eta}_{1}}{dt}\right)\bdot{\bf B}_{0} \right] 
\nonumber \\
 &  &+\; \int_{\bf p} f_{0}\;e\,\vb{\eta}_{1}\bdot\left( {\bf E}_{1} + \frac{{\bf v}_{0}}{c}\btimes{\bf B}_{1}\right) \nonumber \\
 &  &+\; \frac{1}{8\pi} \left( |{\bf E}_{1}|^{2} -\frac{}{} |{\bf B}_{1}|^{2} \right).
 \end{eqnarray}
The gauge-independent Noether equation associated with this gauge-independent quadratic Lagrangian is expressed as $\delta{\cal L}_{2}^{\prime} = \partial\delta{\cal J}_{2}^{\prime}/\partial t + \nabla\bdot\delta\vb{\Gamma}_{2}^{\prime}$, where the Noether fields are
 \begin{eqnarray}
 \delta{\cal J}_{2}^{\prime} & = & \int_{\bf p} f_{0}\;\delta\vb{\eta}_{1}\bdot\left( m\,\frac{d_{0}\vb{\eta}_{1}}{dt} - \frac{e}{2c}\;\vb{\eta}_{1}\btimes{\bf B}_{0}\right) \nonumber \\
  &  &-\; \delta{\bf A}_{1}\bdot\frac{{\bf D}_{1}}{4\pi\,c},
  \end{eqnarray}
  and
  \begin{eqnarray}
 \delta\vb{\Gamma}_{2}^{\prime} & = & \int_{\bf p} {\bf v}_{0}\;f_{0}\;\delta\vb{\eta}_{1}\bdot\left( m\,\frac{d_{0}\vb{\eta}_{1}}{dt} - \frac{e}{2c}\;\vb{\eta}_{1}\btimes{\bf B}_{0}\right) \nonumber \\
  &  &-\; \frac{1}{4\pi} \left( \delta\Phi_{1}\;{\bf D}_{1} \;+\frac{}{} \delta{\bf A}_{1}\btimes{\bf H}_{1} \right).
 \end{eqnarray}
 The energy conservation law \eqref{eq:energy_2} is now expressed in terms of the gauge-independent energy density
 \begin{eqnarray}
 {\cal E}_{2}^{\prime} & = & \int_{\bf p} f_{0} \left( \frac{m}{2}\;\left|\frac{d_{0}\vb{\eta}_{1}}{dt}\right|^{2} \;-\; \frac{e}{c}\,\vb{\eta}_{1}\btimes{\bf v}_{0}\bdot{\bf B}_{1} \right) \nonumber \\
  &  &+\; \frac{1}{8\pi} \left( |{\bf E}_{1}|^{2} \;+\frac{}{} |{\bf B}_{1}|^{2} \right), 
  \end{eqnarray}
 and the gauge-independent energy-density flux
 \begin{eqnarray}
 {\bf S}_{2}^{\prime} & = & \int_{\bf p} {\bf v}_{0}\;f_{0} \frac{d_{0}\vb{\eta}_{1}}{dt}\bdot\left( m\;\frac{d_{0}\vb{\eta}_{1}}{dt} \;-\; \frac{e}{2c}\,\vb{\eta}_{1}\btimes{\bf v}_{0} \right) \nonumber \\
  &  &+\; \frac{{\bf E}_{1}\btimes{\bf H}_{1}}{4\pi}.
 \end{eqnarray}
 The wave-action conservation law $\partial\ov{\cal J}_{2}^{\prime}/\partial t + \nabla\bdot\ov{\vb{\Gamma}}_{2}^{\prime} = 0$, on the other hand, is expressed in terms of the gauge-invariant wave-action density
 \begin{eqnarray}
 \ov{\cal J}_{2}^{\prime} & = & -\; {\rm Im}\left[ \int_{\bf p} f_{0}\;\vb{\eta}_{1}^{*}\bdot\left( m\;\frac{d_{0}\vb{\eta}_{1}}{dt} - \frac{e}{2c}\,\vb{\eta}_{1}\btimes{\bf v}_{0} \right)
\right] \nonumber \\
  &  &+\; {\rm Im}\left(\frac{{\bf A}_{1}^{*}\bdot{\bf D}_{1}}{4\pi\,c}\right),
 \label{eq:J2_prime}
 \end{eqnarray}
 and the gauge-invariant wave-action-density flux
 \begin{eqnarray}
\ov{\vb{\Gamma}}_{2}^{\prime} & = & -\; {\rm Im}\left[ \int_{\bf p} {\bf v}_{0}\;f_{0}\;\vb{\eta}_{1}^{*}\bdot\left( m\;\frac{d_{0}\vb{\eta}_{1}}{dt} - \frac{e}{2c}\,\vb{\eta}_{1}\btimes{\bf v}_{0} \right) \right] \nonumber \\
  &  &+\;  {\rm Im}\left[ \frac{1}{4\pi} \left( \Phi_{1}^{*}\;{\bf D}_{1} \;+\frac{}{} {\bf A}_{1}^{*}\btimes{\bf H}_{1}\right) \right],
 \label{eq:Gamma2_prime}
 \end{eqnarray}
which are identical to expressions derived from the standard Low Lagrangian \cite{BCK_1993}. We note, here, that the gauge invariance of Eqs.~\eqref{eq:J2_prime}-\eqref{eq:Gamma2_prime} follows directly from the gauge-independent first-order Maxwell equations \eqref {eq:Maxwell_DH}.

\section{\label{sec:reduced}Quadratic Lagrangians for reduced Vlasov-Maxwell models}

In this Section, we now look at some applications of the quadratic Lagrangian density \eqref{eq:action_quad} when phase-space transformations are used in the context of dynamical reduction \cite{Brizard_2008}. The guiding-center transformation plays a fundamental role in our understanding of the magnetic confinement of charged particles \cite{Littlejohn_1983,Cary_Brizard}, and serves as an important foundation for the construction of most reduced plasma models. Here, we consider the guiding-center transformation of the quadratic action functional \eqref{eq:action_quad}, from which the variational principles for the linearized gyrokinetic Vlasov-Maxwell equations \eqref{eq:action_quad_gy_nonad}, the linear drift-wave equation \eqref{eq:action_quad_dw}, and the linear gyrokinetic-MHD equations 
\eqref{eq:L_kMHD} are derived.

We begin with the transformation the quadratic action function \eqref{eq:action_quad} to its guiding-center form
\begin{widetext}
\begin{equation}
{\cal L}_{2{\rm gc}} \;=\; \int_{\bf P} \;\left[ \left\{ S_{1{\rm gc}},\frac{}{} F_{0}\right\}_{\rm gc} \left( \frac{1}{2}\,\frac{d_{\rm gc}}{dt}
S_{1{\rm gc}} \;-\; H_{1{\rm gc}}\right) \;-\; \frac{e^{2}\,F_{0}}{2\,mc^{2}}\,|{\bf A}_{1{\rm gc}}|^{2} \right] + \frac{1}{8\pi} \left( |{\bf E}_{1}|^{2} - |{\bf B}_{1}|^{2} \right),
\label{eq:action_quad_gc}
\end{equation}
\end{widetext}
where $F_{0}$ denotes the unperturbed guiding-center Vlasov distribution, $\int_{\bf P} \equiv \int d^{3}P$ includes the guiding-center Jacobian, $d_{\rm gc}/dt$ denotes the (unperturbed) guiding-center Hamiltonian evolution operator, and $\{\;,\;\}_{\rm gc}$ denotes the non-canonical guiding-center Poisson bracket \cite{Littlejohn_1983, Cary_Brizard}. 

In Eq.~\eqref{eq:action_quad_gc}, we also transformed the first-order Hamiltonian $H_{1} \rightarrow H_{1{\rm gc}} = e\,\Phi_{1{\rm gc}} - e\,{\bf A}_{1{\rm gc}}\bdot{\bf v}_{\rm gc}/c \equiv e\,\psi_{1{\rm gc}}$, where 
${\bf v}_{\rm gc} \equiv {\sf T}_{\rm gc}^{-1}{\bf v}$ denotes the guiding-center push-forward of the particle velocity (which includes the guiding-center drift velocity) and the electromagnetic potentials $(\Phi_{1{\rm gc}}, {\bf A}_{1{\rm gc}})$ are evaluated at the particle position ${\bf x} \equiv {\bf X} + \vb{\rho}_{\rm gc}$ expressed in terms of the guiding-center position ${\bf X}$ and the local gyroradius $\vb{\rho}_{\rm gc}$ (which includes higher-order corrections due to magnetic-field nonuniformity \cite{Brizard_2013}). 

In addition, we transformed the first-order generating function $S_{1} \rightarrow S_{1{\rm gc}}$, where the guiding-center generating function $S_{1{\rm gc}} \equiv \langle S_{1{\rm gc}}\rangle + 
\wt{S}_{1{\rm gc}}$ is decomposed into its gyroangle-averaged (nonadiabatic) part $\langle S_{1{\rm gc}}\rangle \equiv S_{1{\rm gy}}$, which defines the first-order gyrocenter generating function $S_{1{\rm gy}}$ \cite{Brizard_1994a}, and its gyroangle-dependent (adiabatic) part $\wt{S}_{1{\rm gc}}$, which satisfies the first-order equation \cite{Brizard_Hahm_2007,Brizard_1989}
\begin{equation}
\frac{d_{\rm gc}}{dt}\,\wt{S}_{1{\rm gc}} \;=\; e\;\wt{\psi}_{1{\rm gc}} \;\;\rightarrow\;\; \wt{S}_{1{\rm gc}} \;=\; \frac{e}{\Omega}\;
\wt{\Psi}_{1{\rm gc}}.
\label{eq:Psi_gc_def}
\end{equation}
We note that only the gyroangle-independent part $S_{1{\rm gy}}$ will appear in the reduced quadratic gyrokinetic Lagrangian density \eqref{eq:action_quad_gy}.

When we insert these decompositions into the guiding-center quadratic action functional \eqref{eq:action_quad_gc}, we obtain the low-frequency gyrocenter Lagrangian density
\begin{eqnarray}
{\cal L}_{2{\rm gy}} & = & \int_{\ov{\bf P}} \left[ \{ S_{1{\rm gy}},\, \ov{F}_{0}\}_{\rm gc} \left( \frac{1}{2}\frac{d_{\rm gc}}{dt}\,
S_{1{\rm gy}} - \langle H_{1{\rm gc}}\rangle \right) \right. \nonumber \\
  &  &\left.-\frac{}{} \ov{F}_{0}\;H_{2{\rm gy}} \right] + \frac{1}{8\pi} \left( |{\bf E}_{1}|^{2} - |{\bf B}_{1}|^{2} \right),
\label{eq:action_quad_gy}
\end{eqnarray}
where the low-frequency perturbed electric field ${\bf E}_{1} = -\;\nabla_{\bot}\Phi_{1}$ is used in the Maxwell part, the operation of gyroangle-averaging was performed in the gyrocenter Vlasov part, with the unperturbed gyrocenter Vlasov distribution $\ov{F}_{0}(\ov{\cal E}, \ov{\mu}, \ov{\bf X})$ depending on the gyrocenter position $\ov{\bf X}$, the gyrocenter magnetic moment $\ov{\mu}$, and the guiding-center kinetic energy $\ov{\cal E} \equiv \ov{H}_{0{\rm gc}}$, and the second-order gyrocenter Hamiltonian is \cite{Brizard_1989}
\begin{equation}
H_{2{\rm gy}} = \frac{e^{2}\left\langle|{\bf A}_{1{\rm gc}}|^{2}\right\rangle}{2\,mc^{2}} - \frac{e^{2}}{2\Omega} \left\langle
\left\{ \wt{\Psi}_{1{\rm gc}}, \wt{\psi}_{1{\rm gc}}\right\}_{\rm gc}\right\rangle.
\label{eq:H2_gy_def}
\end{equation}
We note that the last term in the second-order gyrocenter Hamiltonian \eqref{eq:H2_gy_def} represents the low-frequency ponderomotive Hamiltonian from which the gyrocenter polarization and magnetization effects arise \cite{Brizard_Hahm_2007}. The gyrocenter quadratic action functional \eqref{eq:action_quad_gy} was used to construct the quadratic gyrokinetic free-energy functional \cite{Brizard_1994b}.

We note that the relation between the particle Vlasov distribution $f$ and the gyrocenter Vlasov distribution $\ov{F}$ is expressed in terms of the guiding-center and gyrocenter pull-back operators 
$f \equiv {\sf T}_{\rm gc}({\sf T}_{\rm gy}\,\ov{F})$, which yields \cite{Brizard_1994a}
\begin{eqnarray}
f_{1} & \equiv & {\sf T}_{\rm gc} \left( \ov{F}_{1} \;+\; \frac{e}{\Omega}\,\left\{ \wt{\Psi}_{1{\rm gc}},\frac{}{} \ov{F}_{0}\right\}_{\rm gc} \right. \nonumber \\
 &  &\left.\hspace*{0.5in}+\; \frac{e}{c}\,{\bf A}_{1{\rm gc}}\bdot\left\{ \ov{\bf X} + \ov{\vb{\rho}}_{\rm gc},\; \ov{F}_{0}\right\}_{\rm gc} \right),
\label{eq:fF_1}
\end{eqnarray}
where the first-order gyrocenter Vlasov distribution $\ov{F}_{1}$ is generated by the first-order gyrocenter function $S_{1{\rm gy}}$:
\begin{eqnarray}
\ov{F}_{1} & \equiv & \{ S_{1{\rm gy}}, \ov{F}_{0}\}_{\rm gc} \nonumber \\
 & = & \{S_{1{\rm gy}}, \ov{\cal E}\}_{\rm gc}\pd{\ov{F}_{0}}{\ov{\cal E}} + \frac{c\bhat}{eB_{\|}^{*}}\btimes
\ov{\nabla}\ov{F}_{0}\bdot\ov{\nabla}S_{1{\rm gy}}.
\label{eq:FS_gy}
\end{eqnarray}
We now introduce the nonadiabatic part of the first-order gyrocenter Vlasov distribution \cite{Brizard_1994a}
\begin{eqnarray}
\ov{G}_{1} & \equiv & \ov{F}_{1} \;-\; \langle H_{1{\rm gc}}\rangle\;\pd{\ov{F}_{0}}{\ov{\cal E}}  \nonumber \\
 & = & \left\{ S_{1{\rm gy}},\frac{}{} \ov{F}_{0} \right\}_{\rm gc} \;-\; \frac{d_{\rm gc}\ov{S}_{1{\rm gy}}}{dt}\;
\pd{\ov{F}_{0}}{\ov{\cal E}} \label{eq:G1gy_def}  \\
 & = & \left( \frac{c\bhat}{eB_{\|}^{*}}\btimes\ov{\nabla}\ov{F}_{0}\bdot\ov{\nabla} - \pd{\ov{F}_{0}}{\ov{\cal E}}\pd{}{t}\right)\,S_{1{\rm gy}} \equiv \wh{{\cal Q}}S_{1{\rm gy}},
\nonumber
\end{eqnarray}
where the operator $\wh{\cal Q}$ commutes with $d_{\rm gc}/dt$. With this decomposition, the gyrocenter quadratic Lagrangian density \eqref{eq:action_quad_gy} becomes
\begin{eqnarray}
{\cal L}_{2{\rm gy}} & = & \int_{\ov{\bf P}} \left[ \wh{{\cal Q}}S_{1{\rm gy}} \left(\; \frac{1}{2}\;\frac{d_{\rm gc}}{dt}\,
S_{1{\rm gy}} \;-\; \langle H_{1{\rm gc}}\rangle \;\right) \right. \nonumber \\
 &  &\left.\hspace*{0.3in}-\; \ov{F}_{0}\;\left( H_{2{\rm gy}} \;-\; \frac{1}{2}\;\pd{\langle H_{1{\rm gc}}
\rangle^{2}}{\ov{\cal E}} \right) \right] \nonumber \\
 &  &+\; \frac{1}{8\pi} \left( |\nabla_{\bot}\Phi_{1}|^{2} - |\nabla\btimes{\bf A}_{1}|^{2} \right).
\label{eq:action_quad_gy_nonad}
\end{eqnarray}
The gyrocenter quadratic action functional \eqref{eq:action_quad_gy_nonad} can be used to derive the nonadiabatic gyrokinetic Vlasov-Maxwell equations, which includes the guiding-center and gyrocenter polarizations and magnetizations. 

We will now show that it can also be used to derive the variational formulations for the linear drift-wave equation as well as the linear hybrid gyrokinetic-MHD equations, which describe how the perturbed Vlasov distribution (generated by $S_{1}$) corresponding to an energetic-particle population can be self-consistently linked to a macroscopic plasma mode (described by the ideal MHD fluid displacement $\vb{\xi}_{1}$) in a bulk magnetized plasma.

\subsection{Linear drift-wave equations}

As a first example of the modular property of the variational formulations of reduced plasma models, where different physical effects can be added in modular fashion to an action functional, the nonadiabatic gyrocenter quadratic Lagrangian density \eqref{eq:action_quad_gy_nonad} was previously \cite{Brizard_1996} used to derive the linear drift-wave equation for electrostatic fluctuations $\Phi_{1} = \Phi$ (with ${\bf A}_{1} = 0$) in a cold-ion magnetized plasma (represented by the nonuniform plasma density $n_{0}$ and the uniform magnetic field ${\bf B} = B\,\wh{\sf z}$) with adiabatic electrons (at a uniform temperature $T_{\rm e}$). 

The quadratic drift-wave action functional ${\cal A}_{\rm dw} \equiv \int{\cal L}_{\rm dw}\,d^{3}r dt$ is expressed in terms of the drift-wave Lagrangian density \cite{Brizard_1996}
\begin{eqnarray}
{\cal L}_{\rm dw} & = & \frac{c\wh{\sf z}}{eB}\btimes\nabla n_{0}\bdot\nabla(e\,\psi) \left( \frac{e}{2}\,\pd{\psi }{t} \;-\; e\,\Phi\right) \nonumber \\
 &  &+\; \frac{m_{\rm i}c^{2}\,n_{0}}{2\,B^{2}}\;|\nabla_{\bot}\Phi|^{2} \;+\; \frac{n_{0}\,e^{2}}{2\;T_{\rm e}}\;\Phi^{2} \nonumber \\
 & \equiv & {\bf c}\bdot\nabla_{\bot}\psi\;\left( \frac{1}{2}\,\pd{\psi}{t} \;-\; \Phi\right) \nonumber \\
 &  &+\; \frac{1}{2} \left( b\,|\nabla_{\bot}\Phi|^{2} \;+\frac{}{} a\,\Phi^{2} \right).
\label{eq:action_quad_dw}
\end{eqnarray}
Here, the first term represents the nonadiabatic cold-ion contribution, where the gyrocenter phase-space function $S_{1{\rm gy}} \rightarrow e\psi({\bf x},t)$ is replaced by a scalar field $\psi({\bf x},t)$ in physical space. We note that this additional scalar field contributes to the first-order ion fluid displacement 
\begin{equation}
\vb{\xi}_{1} \;=\; (c\wh{\sf z}/B)\btimes\nabla_{\bot}\psi \;-\; (m_{\rm i}c^{2}/eB^{2})\nabla_{\bot}\Phi,
\label{eq:xi1_dw}
\end{equation}
which is obtained in the cold drift-kinetic limit of the ion gyrocenter displacement $\langle\{ {\bf X} + \vb{\rho}_{\rm gc},\; S_{1{\rm gc}}\}_{\rm gc}\rangle$. The second term $\frac{1}{2}\,b\,|\nabla_{\bot}\Phi|^{2}$ in 
Eq.~\eqref{eq:action_quad_dw}, which arises from the term $-\,\ov{F}_{0}\,H_{2{\rm gy}}$ in Eq.~\eqref{eq:action_quad_gy_nonad}, represents the contribution from the cold-ion gyrocenter polarization (which is much greater than the Maxwell contribution $|\nabla_{\bot}\Phi|^{2}/8\pi$). The third term $\frac{1}{2}\,a\,\Phi^{2}$, which arises from the electron contribution $-\,\frac{1}{2}\,e^{2}\Phi^{2}\,(\partial f_{\rm e}/\partial{\cal E})$ in Eq.~\eqref{eq:action_quad_gy_nonad}, represents the contribution from the adiabatic electrons. We note that all three background-plasma functions $(a, b, {\bf c})$ depend on position through the nonuniform plasma density $n_{0}$ and the vector function ${\bf c}$ is divergenceless in a uniform magnetic field (i.e., $\nabla\bdot{\bf c} = 0$).

The drift-wave variational principle $\delta{\cal A}_{\rm dw} = 0$, based on Eq.~\eqref{eq:action_quad_dw}, yields the coupled equations 
\begin{equation}
\left. \begin{array}{rcl}
\partial\psi/\partial t & = & \Phi \\
a\,\Phi - \nabla_{\bot}\bdot(b\,\nabla_{\bot}\Phi) & = & {\bf c}\bdot\nabla_{\bot}\psi
\end{array} \right\},
\label{eq:dw_psi_phi}
\end{equation}
from which we recover the linear drift-wave equation
\begin{eqnarray}
\pd{}{t}\left[ a\Phi \frac{}{}- \nabla_{\bot}\vb{\cdot}\left(b\nabla_{\bot}\Phi\right) \right] & = & {\bf c}\vb{\cdot}\nabla_{\bot}\pd{\psi}{t} = {\bf c}\vb{\cdot}\nabla_{\bot}\Phi.
\label{eq:ldw}
\end{eqnarray}
We note that the second equation in Eq.~\eqref{eq:dw_psi_phi} can be rewritten in the form of the quasineutrality condition $e\,n_{e1} = e\,n_{i1}$:
\[ a\,\Phi \;=\; -\,\nabla\bdot\left(e\,n_{0}\frac{}{}\vb{\xi}_{1}\right) \;=\; {\bf c}\bdot\nabla\psi \;+\; \nabla_{\bot}\bdot(b\,\nabla_{\bot}\Phi), \]
where the first-order ion fluid displacement \eqref{eq:xi1_dw} was used.

The drift-wave Lagrangian density \eqref{eq:action_quad_dw} can also be used to derive the drift-wave Noether equation $\delta{\cal L}_{\rm dw} = \partial_{t}\delta{\cal J}_{\rm dw} + \nabla\bdot\delta
\vb{\Gamma}_{\rm dw}$, where
\begin{equation}
\left. \begin{array}{rcl}
\delta{\cal J}_{\rm dw} & = & \frac{1}{2}\,\delta\psi\;{\bf c}\bdot\nabla\psi \\
\delta\vb{\Gamma}_{\rm dw} & = & \delta\Phi\;b\,\nabla\Phi - \frac{1}{2} \delta\psi\;{\bf c}\,\partial_{t}\psi
\end{array} \right\},
\end{equation}
 from which the energy-momentum conservation laws for the linear drift-wave equation \eqref{eq:ldw} are derived. For example, the energy conservation law 
$\partial_{t}{\cal E}_{\rm dw} + \nabla\bdot{\bf S}_{\rm dw} = 0$ is expressed in terms of the drift-wave densities
\begin{equation}
\left. \begin{array}{rcl}
{\cal E}_{\rm dw} & = & \Phi\,{\bf c}\bdot\nabla\psi - \frac{1}{2} (b\,|\nabla_{\bot}\Phi|^{2} + a\,\Phi^{2}) \\
{\bf S}_{\rm dw} & = & b\,\nabla_{\bot}\Phi\;\partial_{t}\Phi - \frac{1}{2}\;{\bf c}\;\Phi^{2}
\end{array} \right\}.
\end{equation}
The drift-wave Noether equation was also used to derive the linear drift-wave action conservation law \cite{Brizard_1996}
\begin{equation}
\pd{\ov{\cal J}_{\rm dw}}{t} \;+\; \nabla\bdot\ov{\vb{\Gamma}}_{\rm dw} \;=\; 0,
\label{eq:drift_action}
\end{equation}
where the linear drift-wave action density $\ov{\cal J}_{\rm dw}$ and the linear drift-wave-action-density flux $\ov{\vb{\Gamma}}_{\rm dw}$ are 
\begin{eqnarray}
\ov{\cal J}_{\rm dw} & \equiv & {\rm Im}\left( \frac{1}{2}\,\psi\frac{}{}{\bf c}\bdot\nabla\psi^{*}\right), \\
\ov{\vb{\Gamma}}_{\rm dw} & \equiv & {\rm Im}\left( b\;\Phi\,\nabla\Phi^{*} \;-\; \frac{1}{2}\,{\bf c}\;\psi\,\pd{\psi^{*}}{t}\right).
\end{eqnarray}
The linear drift-wave action conservation law was first derived in {\it ad-hoc} fashion by Mattor and Diamond \cite{Mattor_Diamond_1994} to investigate the role of the drift-wave-action conservation law in drift-wave turbulence propagation. We note that, in the eikonal limit $(\partial/\partial t, \nabla) \rightarrow (-i\,\omega, i{\bf k})$, the drift-wave dispersion relation is 
\[ \omega  \;=\; -\,{\bf k}_{\bot}\bdot{\bf c}/(a + b\,k_{\bot}^{2}) \equiv \omega_{\rm dw}({\bf k}_{\bot}) \]
and the eikonal-averaged drift-wave energy density is $\ov{\cal E}_{\rm dw} = \omega_{\rm dw}\,\ov{\cal J}_{\rm dw}$, where the drift-wave eikonal-averaged action density is $\ov{\cal J}_{\rm dw} = -\,{\bf k}\bdot{\bf c}\,|\wt{\psi}|^{2} \equiv \partial\ov{\cal L}_{\rm dw}/\partial\omega$, while the eikonal-averaged drift-wave energy density is $\ov{\bf S}_{\rm dw} = {\bf v}_{\rm dw}\,\ov{\cal E}_{\rm dw} = \omega_{\rm dw}\,\ov{\vb{\Gamma}}_{\rm dw}$, where the drift-wave group velocity is ${\bf v}_{\rm dw} \equiv \partial\omega_{\rm dw}/\partial{\bf k}$ and the drift-wave eikonal-averaged action density flux is $\ov{\vb{\Gamma}}_{\rm dw} = {\bf v}_{\rm dw}\,\ov{\cal J}_{\rm dw} \equiv -\,\partial\ov{\cal L}_{\rm dw}/\partial{\bf k}$.

\textcolor{red}{Lastly, we need to emphasize again that the drift-wave action conservation law \eqref{eq:drift_action} is no longer valid once additional physics (e.g., the presence of a mean flow) or nonlinear effects (e.g., drift-wave/zonal-flow interactions \cite{Smolyakov_Diamond_1999,Chen_Lin_2000}) are taken into account. On the one hand, additional physics (within the same perturbation order) can always be introduced in the appropriate Lagrangian density in order to derive a generalized form of the wave-action conservation law (see the next modular example). On the other hand, the linear drift-wave action density $\ov{\cal J}_{\rm dw}$ can still be used as a convenient field variable in the ensuing nonlinear-wave discussion.}

\subsection{Linear hybrid gyrokinetic-MHD equations}

Another modular application of the nonadiabatic gyrocenter quadratic action functional \eqref{eq:action_quad_gy_nonad} involves the variational derivation of the standard hybrid gyrokinetic-MHD equations \cite{CWR_1984}. \textcolor{red}{In the presence of an energetic-particle species, the ideal MHD wave-action conservation law is no longer satisfied since the interaction between the linear MHD modes and the perturbations of the energetic-particle distribution must be taken into account. We now show that the introduction of the additional physics associated with the energetic-particle species is done through a modification of the ideal MHD Lagrangian density \cite{Brizard_1992}.}

First, we write the perturbed Hamiltonian for the energetic-particles in the drift-kinetic (dk) limit:
\begin{equation}
\langle H_{1{\rm gc}}\rangle \;\rightarrow\; e\,\left( \Phi_{1} - {\bf A}_{1}\bdot{\bf v}_{\rm gc}/c \right) \;+\; \ov{\mu}\; B_{1\|},
\label{eq:H1gc_dk}
\end{equation}
where ${\bf v}_{\rm gc}$ denotes the guiding-center magnetic-drift velocity. In its simplest version, we use $\Phi_{1} \equiv 0 \equiv A_{1\|}$ (i.e., $E_{1\|} \equiv 0$) and ${\bf A}_{1\bot} \equiv \vb{\xi}_{1}\btimes
{\bf B}_{0}$, where $\vb{\xi}_{1}({\bf x},t)$ denotes the ideal MHD fluid displacement and ${\bf B}_{0} = B_{0}\,\bhat_{0}$ denotes the nonuniform background magnetic field, which yields the first-order Hamiltonian
\begin{eqnarray} 
H_{1{\rm dk}} & = & \ov{\mu}\,\bhat_{0}\bdot\nabla\btimes(\vb{\xi}_{1}\btimes{\bf B}_{0}) \;+\; \vb{\xi}_{1}\bdot \frac{e}{c}\,{\bf v}_{\rm gc}\btimes{\bf B}_{0} 
\nonumber \\
 & = & -\,\ov{\mu}B_{0}\left({\bf I} - \bhat_{0}\bhat_{0}\right):\nabla\vb{\xi}_{1} - m\ov{v}_{\|}^{2}\,\bhat_{0}\bhat_{0}:\nabla\vb{\xi}_{1} \nonumber \\
 & \equiv & -\,\ov{\vb{\Pi}}_{0}:\nabla\vb{\xi}_{1},
\label{eq:H1_dk}
\end{eqnarray}
where $\ov{\vb{\Pi}}_{0} \equiv \ov{\mu} B_{0}\,({\bf I} - \bhat_{0}\bhat_{0}) + m\ov{v}_{\|}^{2}\;\bhat_{0}\bhat_{0}$ and the parallel component $\bhat_{0}\bdot\vb{\xi}_{1}$ is naturally absent from $H_{1{\rm dk}}$. 

By combining the Lagrangian contribution from the nonadiabatic drift-kinetic Lagrangian density \eqref{eq:action_quad_gy} for the energetic particles with the Lagrangian density for the ideal MHD equations \cite{Brizard_1992} (associated with a time-dependent magnetized bulk plasma with mass density $\varrho_{0}$, fluid velocity ${\bf u}_{0}$, plasma pressure $p_{0}$, and magnetic field 
${\bf B}_{0}$), we obtain the quadratic kinetic-MHD Lagrangian density \cite{Brizard_1994a}
\begin{eqnarray} 
{\cal L}_{\rm kMHD} & = & \frac{1}{2} \left( \varrho_{0}\,\left|\frac{d_{0}\vb{\xi}_{1}}{dt}\right|^{2} + \vb{\xi}_{1}\bdot{\bf F}_{1}(\vb{\xi}_{1}) \right) \nonumber \\
 &  &+\; \int_{\ov{\bf p}} \wh{\cal Q}S_{1{\rm dk}} \left( \frac{1}{2}\,\frac{d_{\rm gc}S_{1{\rm dk}}}{dt} - H_{1{\rm dk}}(\vb{\xi}_{1}) \right),
\label{eq:L_kMHD}
\end{eqnarray}
where the self-adjoint operator ${\bf F}_{1} \equiv \nabla\bdot{\sf K}_{1}$ includes a time-dependent contribution from the background bulk plasma \cite{Frieman_Rotenberg,Brizard_1992}, with the first-order dyadic tensor ${\sf K}_{1}$ defined as:
\begin{eqnarray}
{\sf K}_{1}(\vb{\xi}_{1}) & = & \vb{\xi}_{1}\;\varrho_{0}\,\frac{d_{0}{\bf u}_{0}}{dt} - {\bf I} \left( p_{1} \;+\; \frac{1}{4\pi} {\bf B}_{1}\bdot{\bf B}_{0}\right) \nonumber \\
 &  &+\; \frac{1}{4\pi} \left( {\bf B}_{0}\,{\bf B}_{1} \;+\frac{}{} 
{\bf B}_{1}\,{\bf B}_{0}\right).
 \label{eq:K1}
 \end{eqnarray}
Here, the total  bulk-plasma time derivative $d_{0}/dt \equiv \partial/\partial t + {\bf u}_{0}\bdot\nabla$ includes the convective derivative with respect to the time-dependent bulk velocity ${\bf u}_{0}$, with the background plasma equation of motion 
 \[ \varrho_{0}\,\frac{d_{0}{\bf u}_{0}}{dt} \;=\; -\,\nabla p_{0} \;+\; (\nabla\btimes{\bf B}_{0})\btimes\frac{{\bf B}_{0}}{4\pi}, \]
 and the perturbed fields $(\varrho_{1},{\bf u}_{1},p_{1},{\bf B}_{1})$ are defined in terms of the ideal-MHD fluid displacement $\vb{\xi}_{1}$ as
 \begin{equation}
 \left. \begin{array}{rcl}
 \varrho_{1} & \equiv & -\,\nabla\bdot(\varrho_{0}\,\vb{\xi}_{1}) \\
 {\bf u}_{1} & \equiv & d_{0}\vb{\xi}_{1}/dt - \vb{\xi}_{1}\bdot\nabla{\bf u}_{0} \\
 p_{1} & \equiv & -\,\gamma\,p_{0}\,(\nabla\bdot\vb{\xi}_{1}) - \vb{\xi}_{1}\bdot\nabla p_{0} \\
 {\bf B}_{1} & \equiv & \nabla\btimes(\vb{\xi}_{1}\btimes{\bf B}_{0})
 \end{array} \right\}.
 \end{equation}
 The self-adjointness of the operator ${\bf F}_{1}(\vb{\xi}_{1})$ is implied by the identity (see App.~\ref{sec:App_A})
 \begin{equation}
 \vb{\xi}_{1}\bdot{\bf F}_{1}(\delta\vb{\xi}_{1}) \;-\; \delta\vb{\xi}_{1}\bdot{\bf F}_{1}(\vb{\xi}_{1}) \;\equiv\; \nabla\bdot\delta{\bf R}_{2}, 
 \label{eq:self_adjoint}
 \end{equation}
 where the quadratic MHD vector field
\begin{eqnarray}
\delta{\bf R}_{2} & \equiv & \left( p_{1}\,\delta\vb{\xi}_{1} \;-\frac{}{} \delta p_{1}\,\vb{\xi}_{1}\right)  + \frac{{\bf B}_{0}}{4\pi} \left[ (\delta\vb{\xi}_{1}\btimes\vb{\xi}_{1})\bdot\frac{}{} \nabla\btimes{\bf B}_{0}\right] \nonumber 
\\
 &  &+\; \left[ (\vb{\xi}_{1}\btimes{\bf B}_{0})\btimes\frac{\delta{\bf B}_{1}}{4\pi} - (\delta\vb{\xi}_{1}\btimes{\bf B}_{0})\btimes\frac{{\bf B}_{1}}{4\pi} \right]
 \label{eq:delta_R2}
 \end{eqnarray}
involves the reference fields $(p_{0},{\bf B}_{0})$. Variation of the kinetic-MHD action functional with respect to $S_{1{\rm dk}}$ yields the linearized drift-kinetic equation
\begin{equation}
\frac{d_{\rm gc}S_{1{\rm dk}}}{dt} \;=\; H_{1{\rm dk}}(\vb{\xi}_{1}) \;=\; -\,\ov{\vb{\Pi}}_{0}:\nabla\vb{\xi}_{1},
\label{eq:S_dk_eq}
\end{equation}
where we used the fact that the operators $\wh{\cal Q}$ and $d_{\rm gc}/dt \equiv \partial/\partial t + {\bf v}_{\rm gc}\bdot\nabla$ commute. Variation with respect to the ideal-MHD displacement $\vb{\xi}_{1}$ yields the linearized ideal-MHD equation of motion
\begin{equation}
\varrho_{0}\;\frac{d_{0}^{2}\vb{\xi}_{1}}{d t^{2}} \;=\; {\bf F}_{1}(\vb{\xi}_{1}) \;-\; \nabla\bdot{\sf P}_{1},
\label{eq:xi_eq}
\end{equation}
which includes the energetic-particle CGL-like stress tensor
\begin{equation}
{\sf P}_{1} = \int_{\ov{\bf p}} \ov{\vb{\Pi}}_{0}\;\wh{{\cal Q}}S_{1{\rm dk}} \equiv P_{1\bot}\;\left({\bf I} - \bhat_{0}\bhat_{0}\right) \;+\; P_{1\|}\;\bhat_{0}\bhat_{0}.
\label{eq:Pi_dk}
\end{equation}
Once the kinetic-MHD equations \eqref{eq:S_dk_eq}-\eqref{eq:xi_eq} have been derived from the variational principle $\int \delta{\cal L}_{\rm kMHD}\,d^{3}x dt = 0$, we obtain the kinetic-MHD Noether equation 
$\delta{\cal L}_{\rm kMHD} = \partial_{t}\delta{\cal J}_{\rm kMHD} + \nabla\bdot\delta\vb{\Gamma}_{\rm kMHD}$, where the Noether fields are
\begin{equation}
\delta{\cal J}_{\rm kMHD} = \varrho_{0}\,\frac{d_{0}\vb{\xi}_{1}}{dt}\vb{\cdot}\delta\vb{\xi}_{1} + \frac{1}{2}\int_{\ov{\bf p}} \delta S_{1{\rm dk}}\left\{ S_{1{\rm dk}}, \ov{F}_{0}\right\}_{\rm gc}, 
\label{eq:delta_J_kMHD}
\end{equation}
and
\begin{eqnarray}
\delta\vb{\Gamma}_{\rm kMHD} & = & \varrho_{0}{\bf u}_{0}\,\frac{d_{0}\vb{\xi}_{1}}{dt}\bdot\delta\vb{\xi}_{1} \;+\; \int_{\ov{\bf p}} \wh{\cal Q}S_{1{\rm dk}} \;\ov{\vb{\Pi}}_{0}\bdot\delta\vb{\xi}_{1}  \nonumber \\
 &  &+\; \frac{1}{2}\,\delta{\bf R}_{2} \;+\; \frac{1}{2} \int_{\ov{\bf p}} \left( {\bf v}_{\rm gc}\;\delta S_{1{\rm dk}}\left\{ S_{1{\rm dk}}, \ov{F}_{0}\right\}_{\rm gc}  \right. \nonumber \\
 &  &\left.-\; \frac{}{} \delta S_{1{\rm dk}}\,H_{1{\rm dk}}(\vb{\xi}_{1})\;\left\{ \ov{\bf X}, \ov{F}_{0} 
 \right\}_{\rm gc} \right),
  \label{eq:delta_Gamma_kMHD}
\end{eqnarray}
where $\delta{\bf R}_{2}$ is defined in Eq.~\eqref{eq:delta_R2}. 

\subsubsection{Kinetic-MHD energy principle}

Instead of deriving the energy conservation law for the kinetic-MHD equations \eqref{eq:S_dk_eq}-\eqref{eq:xi_eq}, it is customary to derive the standard kinetic-MHD energy principle \cite{CWR_1984,Chen_Zonca_2016} (for a time-independent, stationary background plasma):
\begin{equation}
\omega^{2}\,{\cal I}_{\rm MHD} \;=\; {\cal W}_{\rm MHD} \;+\; {\cal K}_{\rm dk}(\omega),
\label{eq:kMHD_energy}
\end{equation}
which can be directly obtained from Eq.~\eqref{eq:xi_eq} as
\[ 0 \;=\; \int_{\bf x}{\rm Re}\left[\wt{\vb{\xi}}_{1}^{*}\bdot\left(\wt{\bf F}_{1} - \nabla\bdot\wt{\sf P}_{1}(\omega) + \varrho_{0}\omega^{2}\,\wt{\vb{\xi}}_{1}\right)\right], \]
with $(\vb{\xi}_{1},S_{1{\rm dk}}) \equiv (\wt{\vb{\xi}}_{1},\wt{S}_{1{\rm dk}})\,e^{-i\omega t}$. Here, the MHD integrals are the inertia ${\cal I}_{\rm MHD} \equiv \int_{\bf x}\varrho_{0}|\wt{\vb{\xi}}_{1}|^{2}$ and the potential energy ${\cal W}_{\rm MHD} \equiv -\int_{\bf x}{\rm Re}(\wt{\vb{\xi}}_{1}^{*}\bdot\wt{\bf F}_{1})$. The energetic-particle integral ${\cal K}_{\rm dk}(\omega) \equiv \int_{\bf x}{\rm Re}[\wt{\vb{\xi}}_{1}^{*}\bdot
\nabla\bdot\wt{\sf P}_{1}(\omega)]$, on the other hand, is defined as
\begin{eqnarray} 
{\cal K}_{\rm dk}(\omega) & = & -\,\int_{\bf x}{\rm Re}\left[\wt{\sf P}_{1}(\omega):\nabla\wt{\vb{\xi}}_{1}^{*} \right] \nonumber \\
 & = & \int_{({\bf x},\ov{\bf p})}{\rm Re}\left[\wh{\cal Q}(\omega)\wt{S}_{1{\rm dk}}(\omega)\;\wt{H}_{1{\rm dk}}^{*}\right], 
 \end{eqnarray}
 where we have used $\wt{H}_{1{\rm dk}}^{*} = -\,\ov{\vb{\Pi}}_{0}:\nabla\wt{\vb{\xi}}_{1}^{*}$ and we have omitted the surface-integral contribution (since $\ov{F}_{0}$ is expected to vanish at the plasma surface). We note that ${\cal K}_{\rm dk}(\omega)$ is an intricate function of the mode frequency $\omega$, where 
\[ \wh{\cal Q}(\omega) \;\equiv\; i\,\omega\;\pd{\ov{F}_{0}}{\ov{\cal E}} \;+\; \frac{c\bhat_{0}}{eB_{\|}^{*}}\btimes\nabla\ov{F}_{0}\bdot\nabla \]
and $\wt{S}_{1{\rm dk}}(\omega)$ is related to $\wt{H}_{1{\rm dk}} = -\ov{\vb{\Pi}}_{0}:\nabla\wt{\vb{\xi}}_{1}$ through $-i(\omega - \wh{\omega}_{\rm gc})\wt{S}_{1{\rm dk}}(\omega) = \wt{H}_{1{\rm dk}}$, which involves orbital wave-particle resonances, where $\wh{\omega}_{\rm gc} \equiv -i\,{\bf v}_{\rm gc}\bdot\nabla$.

In the absence of an energetic-particle population $(\ov{F}_{0} = 0)$, ideal MHD stability (i.e., $\omega^{2} > 0)$ requires that ${\cal W}_{\rm MHD} > 0$ for all allowable displacements $\wt{\vb{\xi}}_{1}$. In the presence of an energetic-particle population, however, it is clear that the solution of Eq.~\eqref{eq:kMHD_energy} may yield complex-valued frequencies $\omega$, with ${\rm Im}(\omega) > 0$ corresponding to an instability (even if ${\cal W}_{\rm MHD} \geq 0$). The reader is urged to consult the recent review paper by Chen and Zonca \cite{Chen_Zonca_2016} for further details on the linear stability of ideal MHD modes in the presence of an energetic-particle population.

\subsubsection{Wave-action conservation law}

We note that, in general, the nonuniform bulk plasma may also be time-dependent, so that the total energy-momentum of the kinetic-MHD modes are not conserved (i.e., the kinetic-MHD modes may exchange energy-momentum with the bulk plasma). The kinetic-MHD wave-action, however, is exactly conserved: 
\begin{equation}
\pd{\ov{\cal J}_{\rm kMHD}}{t} \;+\; \nabla\bdot\ov{\vb{\Gamma}}_{\rm kMHD} \;=\; 0,
\label{eq:action_kMHD}
\end{equation}
where the kinetic-MHD wave-action densities $(\ov{\cal J}_{\rm kMHD}, \ov{\vb{\Gamma}}_{\rm kMHD})$ can be derived directly from the Noether densities \eqref{eq:delta_J_kMHD}-\eqref{eq:delta_Gamma_kMHD}. The conservation of the total wave-action associated with the interaction of an energetic-particle species with a background bulk plasmas has been investigated in Refs.~\cite{Brizard_flip,Brizard_DC}.

Instead of this Noether derivation, we again proceed directly from the kinetic-MHD equations \eqref{eq:S_dk_eq}-\eqref{eq:xi_eq} to prove that the wave-action conservation law \eqref{eq:action_kMHD} is indeed exact. First, from Eq.~\eqref{eq:xi_eq}, we evaluate
\[ 0 \;=\; {\rm Im}\left[\vb{\xi}_{1}^{*}\bdot\left({\bf F}_{1} - \nabla\bdot{\sf P}_{1} - \varrho_{0}\frac{d_{0}^{2}{\vb{\xi}}_{1}}{dt^{2}}\right)\right], \]
which yields
\begin{eqnarray}
\varrho_{0}\frac{d_{0}\ov{J}_{\rm MHD}}{dt} & \equiv & -\;{\rm Im}\left(\varrho_{0}\,\vb{\xi}_{1}^{*}\bdot\frac{d_{0}^{2}\vb{\xi}_{1}}{dt^{2}} \right) \nonumber \\
 & = & -\,\nabla\bdot\ov{\vb{\Gamma}}_{\rm MHD} \;+\; {\rm Im}\left(\int_{\ov{\bf p}}\wh{\cal Q}S_{1{\rm dk}}\,H_{1{\rm dk}}^{*}\right) \nonumber \\
  &  &+\; \nabla\bdot{\rm Im}\left( \int_{\ov{\bf p}}\wh{\cal Q}S_{1{\rm dk}}\,\vb{\Pi}_{0}\bdot\vb{\xi}_{1}^{*}\right),
 \label{eq:action_kMHD}
 \end{eqnarray}
where the MHD wave-action is defined in terms of the MHD wave-action
\begin{equation}
\ov{J}_{\rm MHD} \;=\; {\rm Im}\left( \vb{\xi}_{1}\bdot\frac{d_{0}\vb{\xi}_{1}^{*}}{dt}\right),
\label{eq:J_MHD}
\end{equation}
and the MHD wave-action flux density
\begin{eqnarray}
\ov{\vb{\Gamma}}_{\rm MHD} & = & {\rm Im}\left[ p_{1}^{*}\;\vb{\xi}_{1} + \left(\vb{\xi}_{1}^{*}\btimes{\bf B}_{0}\right)\btimes\frac{{\bf B}_{1}}{4\pi} \right] \nonumber \\
  &  &+\; \frac{{\bf B}_{0}}{8\pi}\;{\rm Im}\left(\vb{\xi}_{1}\btimes\vb{\xi}_{1}^{*}\right)\bdot\nabla\btimes{\bf B}_{0},
\end{eqnarray}
which is derived in App.~\ref{sec:App_A} as ${\rm Im}(\vb{\xi}_{1}^{*}\bdot{\bf F}_{1}) = \nabla\bdot\ov{\vb{\Gamma}}_{\rm MHD}$.

Next, we note that, using the definition for $\wh{\cal Q}$, we find
\[ \wh{\cal Q}S_{1{\rm dk}}\,H_{1{\rm dk}}^{*} = \{ S_{1{\rm dk}},\; \ov{F}_{0}\}_{\rm gc} \frac{d_{\rm gc}S_{1{\rm dk}}^{*}}{dt} - \left|\frac{d_{\rm gc}S_{1{\rm dk}}}{dt}\right|^{2}\;\pd{\ov{F}_{0}}{\ov{\cal E}}, \]
and, hence, 
\[ {\rm Im}\left(\wh{\cal Q}S_{1{\rm dk}}\,H_{1{\rm dk}}^{*}\right) \;=\; {\rm Im}\left( \{ S_{1{\rm dk}},\; \ov{F}_{0}\}_{\rm gc} \frac{d_{\rm gc}S_{1{\rm dk}}^{*}}{dt}\right). \]
By making use of the definition for $d_{\rm gc}/dt$, as well as the Jacobi property for the guiding-center Poisson bracket $\{\;,\;\}_{\rm gc}$, we find
\begin{equation}
{\rm Im}\left(\wh{\cal Q}S_{1{\rm dk}}\,H_{1{\rm dk}}^{*}\right) \;=\; -\;\frac{d_{\rm gc}\ov{J}_{\rm dk}}{dt} \;-\; \left\{ \ov{\Psi}_{\rm dk},\; \ov{F}_{0}\right\}_{\rm gc} ,
  \label{eq:dJ_dk}
\end{equation}
where we used the unperturbed guiding-center Vlasov equation $d_{\rm gc}\ov{F}_{0}/dt = 0$ for the energetic-particle species, and we defined
\begin{equation}
\left. \begin{array}{rcl}
\ov{J}_{\rm dk} & \equiv & \frac{1}{2}\,{\rm Im}(\{ S_{1{\rm dk}}^{*},\; \ov{F}_{0}\}_{\rm gc}\;S_{1{\rm dk}}) \\
 &  & \\
\ov{\Psi}_{\rm dk} & \equiv & \frac{1}{2}\,{\rm Im}(S_{1{\rm dk}}H_{1{\rm dk}}^{*})
\end{array} \right\}. 
\label{eq:JPsi_def}
\end{equation}
Lastly, using the Poisson-bracket identity
\[ \int_{\ov{\bf p}} \{ f,\; g\}_{\rm gc} \;\equiv\; \nabla\bdot\left( \int_{\ov{\bf p}} f\frac{}{}\{ {\bf X},\; g \}_{\rm gc}\right), \]
which holds for arbitrary functions $(f,g)$, we integrate Eq.~\eqref{eq:dJ_dk} to obtain
\begin{eqnarray*}
{\rm Im}\left(\int_{\ov{\bf p}}\wh{\cal Q}S_{1{\rm dk}}\,H_{1{\rm dk}}^{*}\right) & = & -\,\pd{\ov{\cal J}_{\rm dk}}{t} - \nabla\bdot\left(\int_{\ov{\bf p}}{\bf v}_{\rm gc}\,\ov{J}_{\rm dk} \right)  \\
 &  &-\; \nabla\bdot\left( \int_{\ov{\bf p}}\ov{\Psi}_{\rm dk}\;\{{\bf X},\; \ov{F}_{0}\}_{\rm gc}\right),
 \end{eqnarray*}
where the kinetic wave-action density is defined as
\begin{equation}
\ov{\cal J}_{\rm dk} \equiv \int_{\ov{\bf p}}\ov{J}_{\rm dk} = \frac{1}{2}\; \int_{\ov{\bf p}} {\rm Im}\left(\{ S_{1{\rm dk}}^{*}, \ov{F}_{0}\}_{\rm gc}S_{1{\rm dk}}\right).
\label{eq:J_dk}
\end{equation}
By combining these expressions, we obtain the exact kinetic-MHD wave-action conservation law \eqref{eq:action_kMHD}, where the total wave-action density
\begin{equation}
\ov{\cal J}_{\rm kMHD} \;=\; \ov{\cal J}_{\rm MHD} \;+\; \ov{\cal J}_{\rm dk},
\end{equation}
is the direct sum of the MHD ($\ov{\cal J}_{\rm MHD} \equiv \varrho_{0}\,\ov{J}_{\rm MHD}$) and kinetic components, while the total wave-action density flux
\begin{eqnarray}
\ov{\vb{\Gamma}}_{\rm kMHD} & = & {\bf u}_{0}\,\ov{\cal J}_{\rm MHD} \;+\; \ov{\vb{\Gamma}}_{\rm MHD} \nonumber \\
  &  &+\; \int_{\ov{\bf p}}\left( {\bf v}_{\rm gc}\,\ov{J}_{\rm dk} \;-\frac{}{} \ov{\Psi}_{\rm dk}\;\{{\bf X},\; 
\ov{F}_{0}\}_{\rm gc}\right) \nonumber \\
  &  &+\; \int_{\ov{\bf p}} {\rm Im}\left(\wh{\cal Q}S_{1{\rm dk}}^{*}\,\vb{\Pi}_{0}\bdot\vb{\xi}_{1}\right),
 \end{eqnarray}
 where $(\ov{J}_{\rm dk}, \ov{\Psi}_{\rm dk})$ are defined in Eq.~\eqref{eq:JPsi_def}, is the sum of MHD, kinetic, and kinetic-MHD coupling components. We note that, while the ideal MHD wave action \eqref{eq:J_MHD} is positive, the sign of the kinetic wave action \eqref{eq:J_dk} is indefinite. Hence, when an energetic-particle population supports a negative-energy wave \cite{Brizard_flip}, the total ideal MHD and kinetic wave actions 
 $\int_{\bf x}{\cal J}_{\rm MHD}$ and $\int_{\bf x}{\cal J}_{\rm dk}$ may grow separately while keeping their sum $\int_{\bf x}({\cal J}_{\rm MHD} + {\cal J}_{\rm dk})$ constant. 
 
 \textcolor{red}{Lastly, we note that the ideal MHD wave-action conservation law $\partial\ov{\cal J}_{\rm MHD}/\partial t + \nabla\bdot({\bf u}_{0}\,\ov{\cal J}_{\rm MHD} + \ov{\vb{\Gamma}}_{\rm MHD} ) \neq 0$ is no longer conserved in the presence of an 
 energetic-particle species. Through the proper Lagrangian description of the additional physics associated with the perturbed energetic-particle distribution, however, a generalized kinetic-MHD wave-action conservation law \eqref{eq:action_kMHD} was derived (by Noether method) directly from the kinetic-MHD Lagrangian density \eqref{eq:L_kMHD}.}

\section{\label{sec:cubic}Third-order Lagrangian density}

We now move on to include nonlinear effects into the perturbed Vlasov-Maxwell equations by including third-order nonlinearities in the perturbed Vlasov-Maxwell action functional. The perturbative action functional \eqref{eq:action_epsilon} yields the third-order Lagrangian density
\begin{eqnarray}
{\cal L}_{3} & = & \frac{1}{4\pi} \left( {\bf E}_{1}\bdot{\bf E}_{2} \;-\frac{}{} {\bf B}_{1}\bdot{\bf B}_{2} \right) \;-\;\int_{\bf p} \frac{e^{2}f_{0}}{mc^{2}}\,{\bf A}_{1}\bdot{\bf A}_{2} \nonumber \\
  &  &+\; \frac{1}{3}\int_{\bf p} f_{2} \left( \frac{d_{0}S_{1}}{dt} - H_{1}\right) \nonumber \\
  &  &+\; \frac{2}{3}\int_{\bf p}\,f_{1} \left( \frac{d_{0}S_{2}}{dt} - H_{2} + \frac{1}{2} \{S_{1},\; H_{1}\} \right) \nonumber \\
  &  &+\; \frac{1}{3}\int_{\bf p} f_{0} \left( \{ S_{1},\; H_{2}\} \;+\frac{}{} 2\{ S_{2},\; H_{1}\} \right),
\label{eq:cubic_action}
\end{eqnarray}
where $f_{1} = \{ S_{1}, f_{0}\}$ and $f_{2} = \{ S_{2}, f_{0}\} + \frac{1}{2} \{ S_{1}, f_{1}\}$, and the contributions from the third-order Hamiltonian associated with $S_{1}$ and $S_{2}$ (as well as $f_{1}$ and $f_{2}$) appear explicitly. We note, here, that the third-order Lagrangian \eqref{eq:cubic_action} does not simply involve terms that are cubic in the first-order fields $(S_{1},\Phi_{1},{\bf A}_{1})$, but also include terms involving the second-order ponderomotive fields $(S_{2},\Phi_{2},{\bf A}_{2})$. This ponderomotive dependence is in contrast to traditional third-order action functionals, which are always cubic in first-order fields. For example, see the early work of Boyd \& Turner \cite{Boyd_Turner_1972}  for the Vlasov-Maxwell equations, the work of Brizard \& Kaufman \cite{Brizard_MR} for the Manley-Rowe relations describing stimulated Raman scattering in an unmagnetized background plasma, and the more recent works of Pfirsch \& Sudan \cite{Pfirsch_Sudan_1993} and Hirota \cite{Hirota_2011} for the ideal MHD equations. 

\subsection{Gauge-invariant formulation}

Variations of the third-order Lagrangian \eqref{eq:cubic_action} with respect to $(S_{2}, \Phi_{2}, {\bf A}_{2})$ yield the first-order Vlasov-Maxwell equations \eqref{eq:S1_eq}-\eqref{eq:curl_B1}. Variation with respect to $S_{1}$ yields the second-order Vlasov equation
\begin{eqnarray}
0 & = & \frac{d_{0}f_{2}}{dt} \;+\; \{ f_{1},\; H_{1}\} + \{ f_{0},\; H_{2}\} \nonumber \\
  & = & \left\{ \left(\frac{d_{0}S_{2}}{dt} \;-\; H_{2} \;+\; \frac{1}{2} \{S_{1},\; H_{1}\}\right),\; f_{0} \right\} \nonumber \\
  &  &+\; \frac{1}{2}\;\left\{ \left( \frac{d_{0}S_{1}}{dt} \;-\; H_{1}\right),\; f_{1} \right\},
\end{eqnarray}
which yields the second-order equation
\begin{equation}
\frac{d_{0}S_{2}}{dt} \;=\; H_{2} \;-\; \frac{1}{2} \{S_{1},\; H_{1}\}, 
\label{eq:S2_dot} 
\end{equation}
provided Eq.~\eqref{eq:S1_eq} holds. Variations with respect to $(\Phi_{1}, {\bf A}_{1})$, on the other hand, yield the second-order Maxwell equations
\begin{eqnarray}
\nabla\bdot{\bf E}_{2} & = & 4\pi\,\int_{\bf p} e\;f_{2}, \label{eq:div_E2} \\
\nabla\btimes{\bf B}_{2} - \frac{1}{c}\pd{{\bf E}_{2}}{t} & = & \frac{4\pi}{c}\int_{\bf p} e\left[{\bf v}_{0}\,f_{2} \frac{}{} \right. \nonumber \\
 &  &\left.- \frac{e}{mc}\left({\bf A}_{2}\,f_{0} \;+\frac{}{} {\bf A}_{1}\,f_{1}\right) \right] 
\label{eq:curl_B2}.
 \end{eqnarray}
 While the second-order Vlasov-Maxwell equations \eqref{eq:S2_dot}-\eqref{eq:curl_B2} are gauge-dependent, they are invariant under the second-order gauge transformation
 \begin{equation}
\left. \begin{array}{rcl}
 S_{2} & \rightarrow & S_{2} - (e/c)\,\chi_{2} + \frac{1}{2}\,\{ S_{1}, (e/c)\,\chi_{1}\} \\
 \Phi_{2} &  \rightarrow &  \Phi_{2} - c^{-1}\,\partial\chi_{2}/\partial t \\
 {\bf A}_{2} &  \rightarrow &  {\bf A}_{2} + \nabla\chi_{2}
 \end{array} \right\},
 \label{eq:gauge_2}
 \end{equation}
as well as the first-order gauge transformations \eqref{eq:gauge_1}-\eqref{eq:gauge_fh1}, with
\begin{eqnarray*}
f_{2} & \rightarrow & f_{2} - \frac{e}{c}\,\{ \chi_{2},\; f_{0}\} - \frac{e}{c}\,\left\{ \chi_{1},\frac{}{} f_{1} - \frac{e}{2c}\{ \chi_{1},\; f_{0}\} \right\} , \\
H_{2} & \rightarrow & H_{2} - \frac{e}{c}\,\frac{d_{0}\chi_{2}}{dt} \;+\; \frac{e^{2}\nabla\chi_{1}}{mc^{2}}\bdot\left({\bf A}_{1} + \frac{1}{2}\,\nabla\chi_{1}\right).
\end{eqnarray*}
 
 \subsection{Gauge-independent formulation}
 
 A gauge-independent formulation can also be derived as follows. First, we note that Eq.~\eqref{eq:S2_dot} can be written as
\begin{equation}
\frac{d_{0}S_{2}^{\prime}}{dt} \;=\; e \left( \Phi_{2} - \frac{{\bf v}_{0}}{c}\bdot{\bf A}_{2} \right) \;+\; K_{2}, 
 \label{eq:S2prime_dot}
 \end{equation}
 where the second-order (gauge-independent) ponderomotive Hamiltonian is
\begin{equation}
K_{2} \;\equiv\; -\; \frac{e}{2}\vb{\eta}_{1}\bdot\left({\bf E}_{1} + \frac{{\bf v}_{0}}{c}\btimes{\bf B}_{1}\right),
\label{eq:K2_def}
 \end{equation}
 and we have defined the new scalar function
\begin{equation}
S_{2}^{\prime} \;\equiv\; S_{2} \;+\; \frac{e}{2c}{\bf A}_{1}\bdot\vb{\eta}_{1} \;\equiv\; S_{2} \;+\; \sigma_{2},
\end{equation}
which implies that Eq.~\eqref{eq:S2prime_dot} is invariant under the second-order gauge transformation $(\Phi_{2}, {\bf A}_{2}) \rightarrow (\Phi_{2} - c^{-1}\partial\chi_{2}/\partial t, {\bf A}_{2} + \nabla\chi_{2})$ provided $S_{2}^{\prime}$ transforms as $S_{2}^{\prime} \rightarrow S_{2}^{\prime} - (e/c)\,\chi_{2}$, while it is independent of the first-order gauge field $\chi_{1}$. 

Next, we now introduce the gauge-independent second-order displacement $\vb{\eta}_{2} \equiv \partial S_{2}^{\prime}/\partial{\bf p}$, which yields the gauge-invariant second-order velocity
 \begin{equation}
{\bf u}_{2} \;\equiv\; \frac{d_{0}\vb{\eta}_{2}}{dt} \;-\; \vb{\eta}_{2}\bdot\nabla{\bf v}_{0} \;=\; \pd{K_{2}}{\bf p} \;-\; \frac{1}{m} \left( \nabla S_{2}^{\prime} + \frac{e}{c}\,{\bf A}_{2} \right),
 \end{equation}
and the gauge-independent second-order equation of motion
 \begin{eqnarray}
 \frac{d_{0}}{dt}\left( \frac{d_{0}\vb{\eta}_{2}}{dt} - \pd{K_{2}}{\bf p}\right) & = & e\left({\bf E}_{2} + \frac{{\bf v}_{0}}{c}\btimes{\bf B}_{2}\right) + \{ {\bf v}_{0}, K_{2}\} \nonumber \\
  &  &+\; \left( \frac{d_{0}\vb{\eta}_{2}}{dt} - \pd{K_{2}}{\bf p}\right)\btimes\frac{e{\bf B}_{0}}{mc}.
  \end{eqnarray}
The gauge-independent second-order Maxwell equations
 \begin{eqnarray}
 0 & = & \nabla\bdot( {\bf E}_{2} + 4\pi\,\mathbb{P}_{2}) \equiv \nabla\bdot{\bf D}_{2}, 
 \label{eq:div_D2} \\
 0 & = & \nabla\btimes({\bf B}_{2} - 4\pi\,\mathbb{M}_{2}) - \frac{1}{c}\pd{}{t}({\bf E}_{2}+ 4\pi\,\mathbb{P}_{2}) \nonumber \\
  & \equiv & \nabla\btimes{\bf H}_{2} \;-\; \frac{1}{c}\;\pd{{\bf D}_{2}}{t},
  \label{eq:curl_H2}
 \end{eqnarray}
are expressed in terms of the second-order polarization and magnetization
\begin{eqnarray} 
\mathbb{P}_{2} & \equiv & \int_{\bf p} \left( e\,f_{0}\;\vb{\eta}_{2} \;+\; \frac{e}{2}\,f_{1}^{\prime}\;\vb{\eta}_{1}\right), \label{eq:P2_def} \\
\mathbb{M}_{2} & \equiv & \int_{\bf p} \left( e\,f_{0}\;\vb{\eta}_{2} \;+\; \frac{e}{2}\,f_{1}^{\prime}\;\vb{\eta}_{1}\right)\btimes\frac{{\bf v}_{0}}{c}, \label{eq:M2_def}
 \end{eqnarray}
where $f_{1}^{\prime} \equiv -\,m{\bf u}_{1}\bdot\partial f_{0}/\partial{\bf p} -\,\vb{\eta}_{1}\bdot\nabla f_{0}$ is the gauge-independent first-order Vlasov distribution. Once again, from Eqs.~\eqref{eq:div_D2}-\eqref{eq:curl_H2}, we see that the second-order perturbed charge and current densities are expressed in terms of second-order polarization and magnetization effects. 

Lastly, we note that these second-order equations can be derived from the gauge-independent third-order Lagrangian density
\begin{widetext}
 \begin{eqnarray}
 {\cal L}_{3}^{\prime} & = & \frac{1}{4\pi} \left( {\bf E}_{1}\vb{\cdot}{\bf E}_{2} - {\bf B}_{1}\vb{\cdot}{\bf B}_{2}\right) + \int_{\bf p} f_{0} \left[ \vb{\eta}_{1}\vb{\cdot}\left( {\bf F}_{2} - \nabla K_{2}\right) \;+\;  
 m{\bf u}_{1}\vb{\cdot}\left({\bf u}_{2} - \pd{K_{2}}{\bf p}\right) + \vb{\eta}_{2}\vb{\cdot}\left({\bf F}_{1} - \vb{\eta}_{1}\vb{\cdot}\nabla\nabla H_{0}\right) \right] \nonumber \\
   &  &-\; \frac{1}{6}\int_{\rm p} f_{1}^{\prime}\;\vb{\eta}_{1}\bdot\left(m \frac{d_{0}^{2}\vb{\eta}_{1}}{dt^{2}} - \frac{d_{0}\vb{\eta}_{1}}{dt}\btimes\frac{e}{c}{\bf B}_{0}\right),
   \label{eq:cubic_Lag_free}
  \end{eqnarray}
\end{widetext}
where the gauge-independent $n$th-order perturbed fields $({\bf u}_{n},{\bf F}_{n})$ are the velocities ${\bf u}_{n} \equiv d_{0}\vb{\eta}_{n}/dt - \vb{\eta}_{n}\bdot\nabla{\bf v}_{0}$ and the total forces ${\bf F}_{n} \equiv e\,{\bf E}_{n} + (e/c)\,{\bf v}_{0}\btimes{\bf B}_{n}$. We note that, once again, the gauge-independent third-order Lagrangian density \eqref{eq:cubic_Lag_free} involves terms that are cubic in the first-order fields 
$(\vb{\eta}_{1},{\bf E}_{1},{\bf B}_{1})$ as well as terms that involve the second-order ponderomotive fields $(\vb{\eta}_{2},{\bf E}_{2},{\bf B}_{2})$. Applications of the third-order Lagrangian density \eqref{eq:cubic_action} [or Eq.~\eqref{eq:cubic_Lag_free}] will be explored in future work.

\section{\label{sec:pol_mag}Perturbed Vlasov-Maxwell Polarization and Magnetization}

Before summarizing our work, we note that, by combining the second-order polarization and magnetization \eqref{eq:P2_def}-\eqref{eq:M2_def} with the first-order expressions \eqref{eq:PM_1}, we obtain expressions for the perturbed Vlasov-Maxwell polarization and magnetization that are exact to all orders. First, we find the perturbed Vlasov-Maxwell polarization
\begin{eqnarray}
\mathbb{P} & = & \int_{\bf p} e\left[ \epsilon\,\vb{\eta}_{1}\,f_{0} + \epsilon^{2}\,\left( \vb{\eta}_{2}\,f_{0} + \frac{1}{2}\,\vb{\eta}_{1}\,f_{1}^{\prime} \right) + \cdots \right] \nonumber \\
   & \equiv & \int_{0}^{\epsilon} \left( \int_{\bf p} e\,f\;\frac{d{\bf x}}{d\sigma}\right) d\sigma \;\equiv\; \int_{0}^{\epsilon} \mathbb{P}_{\sigma}\;d\sigma ,
\label{eq:Pol}
\end{eqnarray}
where $d{\bf x}/d\sigma \equiv \{{\bf x}, S\}$, with $f = f_{0} + \sigma\{ S_{1}, f_{0}\} + \cdots$ and $S = S_{1} + 2\sigma\,S_{2} + \cdots$. If we return to Eq.~\eqref{eq:div_E_sigma}, we easily recover
\begin{eqnarray} 
\nabla\bdot\pd{\bf E}{\sigma} & = & 4\pi\int_{\bf p} e\,\pd{f}{\sigma} \;=\; 4\pi \int_{\bf p} e\,\{ S,\; f \} \nonumber \\
   & = & -\;\nabla\bdot\left(4\pi \int_{\bf p} e\,f\;\{{\bf x},\; S\} \right) \nonumber \\
   & \equiv & -\;\nabla\bdot\left(4\pi\frac{}{}
\mathbb{P}_{\sigma}\right),
\label{eq:div_E_sigma_final}
\end{eqnarray}
from which we obtain $\nabla\bdot({\bf E} - {\bf E}_{0}) = -\;4\pi\;\nabla\bdot\mathbb{P}$, i.e.,
\begin{equation}
\nabla\bdot{\bf D} \;=\; \nabla\bdot{\bf E}_{0} \;\equiv\; 4\pi\,\rho_{0}.
\label{eq:div_D_E0}
\end{equation}
Hence, we conclude that the perturbed charge density can be expressed as a perturbed polarization charge density at all orders in Vlasov-Maxwell perturbation theory.

The perturbed Vlasov-Maxwell magnetization, on the other hand, is expressed as
\begin{eqnarray}
\mathbb{M} & \equiv & \int_{\bf p} \left[ \epsilon\,\vb{\eta}_{1}\,f_{0} + \epsilon^{2}\,\left( \vb{\eta}_{2}\,f_{0} + \frac{1}{2}\,\vb{\eta}_{1}\,f_{1}^{\prime} \right) + \cdots \right]\btimes\frac{e{\bf v}_{0}}{c} \nonumber \\
 & \equiv & \int_{0}^{\epsilon} \left[\int_{\bf p}  f\;\frac{e}{c}\,\left(\frac{d{\bf x}}{d\sigma}\btimes\frac{d{\bf x}}{dt}\right)\right]d\sigma \equiv \int_{0}^{\epsilon} \mathbb{M}_{\sigma}\;d\sigma,
\label{eq:Mag}
\end{eqnarray}
where $d{\bf x}/dt \equiv \{{\bf x}, H\} = ({\bf p} - e{\bf A}/c)/m = {\bf v}_{0} - \sigma\,e{\bf A}_{1}/mc + \cdots$. If we return to Eq.~\eqref{eq:curl_B_sigma}, we find
\begin{eqnarray}
\nabla\btimes\pd{\bf B}{\sigma} - \frac{1}{c}\pd{}{t}\left(\pd{\bf E}{\sigma}\right) & = & 4\pi\,\int_{\bf p} \frac{e}{c}\,\pd{}{\sigma}\left(f\;\frac{d{\bf x}}{dt}\right) \nonumber \\
 & = & -\;\nabla\bdot\left( 4\pi\,\int_{\bf p}\frac{e}{c}f\;\frac{d{\bf x}}{d\sigma}\,\frac{d{\bf x}}{dt}\right) \nonumber \\
  &  &+\; 4\pi\,\int_{\bf p}f\;\frac{e}{c}\,\frac{d^{2}{\bf x}}{d\sigma\,dt},
 \label{eq:curl_B_sigma_2}
\end{eqnarray}
where we have used $\partial f/\partial\sigma = -\,\{f,\; S\}$, we have used the identity
\[ \int_{\bf p} \left\{ g,\; S \right\} \;=\; \nabla\bdot\left( \int_{\bf p} \frac{d{\bf x}}{d\sigma}\; g \right), \]
 and
\[ \frac{d^{2}{\bf x}}{d\sigma\,dt} \;\equiv\; \pd{}{\sigma}\left(\frac{d{\bf x}}{dt}\right) \;+\; \left\{ \frac{d{\bf x}}{dt},\; S \right\}. \]
Next, using Eq.~\eqref{eq:Pol} and the same identities, we introduce
\begin{eqnarray*} 
\pd{\mathbb{P}_{\sigma}}{t} & = & \int_{\bf p} e\,\pd{}{t}\left(f\;\frac{d{\bf x}}{d\sigma} \right) \nonumber \\
  & = & -\;\nabla\bdot\left( \int_{\bf p}e\,f\;\frac{d{\bf x}}{dt}\,\frac{d{\bf x}}{d\sigma}\right) \;+\; \int_{\bf p}e\,f\;\frac{d^{2}{\bf x}}{d\sigma\,dt},
 \end{eqnarray*}
which is substituted into Eq.~\eqref{eq:curl_B_sigma_2} to yield
\begin{eqnarray}
&  &\nabla\btimes\pd{\bf B}{\sigma} - \frac{1}{c}\pd{}{t}\left(\pd{\bf E}{\sigma}\right) \nonumber \\
 & = & \frac{4\pi}{c}\;\pd{\mathbb{P}_{\sigma}}{t} \;+\; \nabla\bdot\left[ 4\pi\,\int_{\bf p}\frac{e}{c}f\;\left( \frac{d{\bf x}}{dt}\,\frac{d{\bf x}}{d\sigma}
\;-\; \frac{d{\bf x}}{d\sigma}\,\frac{d{\bf x}}{dt}\right) \right] \nonumber \\
 & \equiv & \frac{4\pi}{c}\;\pd{\mathbb{P}_{\sigma}}{t} \;+\; 4\pi\;\nabla\btimes\mathbb{M}_{\sigma},
 \label{eq:curl_B_sigma_final}
\end{eqnarray}
where $\mathbb{M}_{\sigma}$ is defined in Eq.~\eqref{eq:Mag}. We can, once again, conclude that the perturbed current density can be expressed as a perturbed polarization and magnetization current densities, i.e.,
\begin{equation}
\nabla\btimes{\bf H} \;-\; \frac{1}{c}\,\pd{\bf D}{t} \;=\; \nabla\btimes{\bf B}_{0} \;-\; \frac{1}{c}\,\pd{{\bf E}_{0}}{t} \equiv \frac{4\pi}{c}\,{\bf J}_{0}, 
\label{eq:curl_H_B0}
\end{equation}
at all orders in Vlasov-Maxwell perturbation theory.

\section{\label{sec:summary}Summary}

The perturbative variational formulation \eqref{eq:action_epsilon} of the Vlasov-Maxwell equations has been presented, based on a geometric interpretation of the Lie-transform perturbation analysis \eqref{eq:Sh_constraint}. From the second-order and third-order variational principles \eqref{eq:action_quad} and \eqref{eq:cubic_action}, we derived first-order and second-order Vlasov-Maxwell equations in both gauge-invariant and gauge-independent forms. In the gauge-independent forms, we extracted explicit expressions for the perturbed Vlasov-Maxwell polarization and magnetization \eqref{eq:Pol} and \eqref{eq:Mag}.

From the quadratic variational principle for the linearized Vlasov-Maxwell equations, we derived variational principles for the linear drift-wave equation and the linearized kinetic-MHD equations, from which wave-action conservation laws were derived.

\appendix

\section{\label{sec:App_A}Self-adjointness Property}

In this Appendix, we prove the identity \eqref{eq:self_adjoint}, where ${\bf F}_{1}(\delta\vb{\xi}_{1}) \equiv \delta{\bf F}_{1} = \nabla\bdot\delta{\sf K}_{1}$ is expressed in terms of Eq.~\eqref{eq:K1} as
\begin{eqnarray}
\delta{\sf K}_{1} & = & \delta\vb{\xi}_{1}\;\varrho_{0}\,\frac{d_{0}{\bf u}_{0}}{dt} - {\bf I} \left( \delta p_{1} \;+\; \frac{1}{4\pi} \delta{\bf B}_{1}\bdot{\bf B}_{0}\right) \nonumber \\
 &  &+\; \frac{1}{4\pi} \left( {\bf B}_{0}\,\delta{\bf B}_{1} \;+\frac{}{} \delta{\bf B}_{1}\,{\bf B}_{0}\right),
 \label{eq:delta_K1}
 \end{eqnarray}
where $\delta p_{1} \equiv -\,\gamma p_{0}\,(\nabla\bdot\delta\vb{\xi}_{1}) - \delta\vb{\xi}_{1}\bdot\nabla p_{0}$ and $\delta{\bf B}_{1} \equiv\nabla\btimes(\delta\vb{\xi}_{1}\btimes{\bf B}_{0})$. We begin with
\begin{equation} 
\vb{\xi}_{1}\bdot\delta{\bf F}_{1} \;=\; \nabla\bdot\left(\delta{\sf K}_{1}\bdot\vb{\xi}_{1}\right) \;-\; \delta{\sf K}_{1}^{\top}:\nabla\vb{\xi}_{1},
\end{equation}
so that Eq.~\eqref{eq:self_adjoint} becomes
 \begin{eqnarray}
  \vb{\xi}_{1}\bdot\delta{\bf F}_{1} - \delta\vb{\xi}_{1}\bdot{\bf F}_{1} & = & \nabla\bdot\left( \delta{\sf K}_{1}\bdot\vb{\xi}_{1} - {\sf K}_{1}\bdot\delta\vb{\xi}_{1}\right) \nonumber \\
 &  &+\; {\sf K}_{1}^{\top}:\nabla\delta\vb{\xi}_{1} - \delta{\sf K}_{1}^{\top}:\nabla\vb{\xi}_{1},
\label{eq:id_def} 
 \end{eqnarray}
 where we now have to show that $ {\sf K}_{1}^{\top}:\nabla\delta\vb{\xi}_{1} - \delta{\sf K}_{1}^{\top}:\nabla\vb{\xi}_{1}$ can be written as a divergence. Here, $\delta{\sf K}_{1}^{\top}$ denotes the transpose of 
 $\delta{\sf K}_{1}$ so that
 \begin{widetext}
\[ \delta{\sf K}_{1}^{\top}:\nabla\vb{\xi}_{1} = \varrho_{0}\,\frac{d_{0}{\bf u}_{0}}{dt}\bdot\left(\delta\vb{\xi}_{1}\bdot\nabla\vb{\xi}_{1}\right) - \left( \delta p_{1} + \frac{1}{4\pi} \delta{\bf B}_{1}\bdot{\bf B}_{0}\right) \nabla\bdot\vb{\xi}_{1} + \frac{1}{4\pi} \left( {\bf B}_{0}\bdot\nabla\vb{\xi}_{1}\bdot\delta{\bf B}_{1} \;+\frac{}{} \delta{\bf B}_{1}\bdot\nabla\vb{\xi}_{1}\bdot{\bf B}_{0}\right). \]
 After several manipulations, we find
 \begin{eqnarray*}
 \delta{\sf K}_{1}^{\top}:\nabla\vb{\xi}_{1} & = & \gamma p_{0} (\nabla\bdot\vb{\xi}_{1})\,(\nabla\bdot\delta\vb{\xi}_{1}) + \delta{\bf B}_{1}\bdot \frac{{\bf B}_{1}}{4\pi} - \vb{\xi}_{1}\delta\vb{\xi}_{1}:\nabla\nabla\left( p_{0} + \frac{B_{0}^{2}}{8\pi}\right) + \left[ \vb{\xi}_{1}\bdot\nabla p_{0} \left( \nabla\bdot\delta\vb{\xi}_{1}\right) \;+\frac{}{} \delta\vb{\xi}_{1}\bdot\nabla p_{0} \left( \nabla\bdot\vb{\xi}_{1}\right) \right] \\
   &  &-\; \frac{1}{4\pi} \left(\vb{\xi}_{1}\bdot\nabla{\bf B}_{0}\right)\bdot\left(\delta\vb{\xi}_{1}\bdot\nabla{\bf B}_{0}\right) \;+\; \frac{1}{4\pi} \left[ \left(\vb{\xi}_{1}\bdot\nabla{\bf B}_{0}\right)\bdot\left({\bf B}_{0}\bdot\nabla\delta\vb{\xi}_{1}\right) \;+\frac{}{} \left(\delta\vb{\xi}_{1}\bdot\nabla{\bf B}_{0}\right)\bdot 
     \left({\bf B}_{0}\bdot\nabla\vb{\xi}_{1}\right)\right] \\
       &  &+\; \nabla\bdot\left[ \frac{\delta{\bf B}_{1}}{4\pi} \left(\vb{\xi}_{1}\bdot{\bf B}_{0}\right) \;+\; \delta\vb{\xi}_{1}\left(\vb{\xi}_{1}\bdot\varrho_{0}\,\frac{d_{0}{\bf u}_{0}}{dt}\right) \;-\frac{}{} \frac{{\bf B}_{0}}{4\pi} \left(\delta\vb{\xi}_{1}\bdot\nabla{\bf B}_{0}\bdot\vb{\xi}_{1}\right) \frac{}{} \right],
 \end{eqnarray*}
 where all the terms outside of the divergence terms are explicitly symmetric with respect to $\vb{\xi}_{1}$ and $\delta\vb{\xi}_{1}$. Hence, we easily find that
\[ {\sf K}_{1}^{\top}:\nabla\delta\vb{\xi}_{1} - \delta{\sf K}_{1}^{\top}:\nabla\vb{\xi}_{1} \;=\; \nabla\bdot\left[ \frac{{\bf B}_{1}}{4\pi} \left(\delta\vb{\xi}_{1}\bdot{\bf B}_{0}\right) - \frac{\delta{\bf B}_{1}}{4\pi} 
\left(\vb{\xi}_{1}\bdot{\bf B}_{0}\right) \;-\; \left(\vb{\xi}_{1}\btimes\delta\vb{\xi}_{1}\right)\btimes\varrho_{0}\,\frac{d_{0}{\bf u}_{0}}{dt} \;-\; \frac{{\bf B}_{0}}{4\pi} \left(\vb{\xi}_{1}\btimes\delta\vb{\xi}_{1}\right)\bdot\nabla\btimes{\bf B}_{0} \frac{}{} \right]. \]
If we now substitute this expression, with
\begin{eqnarray*} 
\delta{\sf K}_{1}\bdot\vb{\xi}_{1} - {\sf K}_{1}\bdot\delta\vb{\xi}_{1} & = & \left(\vb{\xi}_{1}\btimes\delta\vb{\xi}_{1}\right)\btimes\varrho_{0}\,\frac{d_{0}{\bf u}_{0}}{dt} \;+\; \left( p_{1}\,\delta\vb{\xi}_{1} \;-\frac{}{}
\delta p_{1}\;\vb{\xi}_{1}\right) \;+\; \frac{1}{4\pi} \left[ (\vb{\xi}_{1}\btimes{\bf B}_{0})\btimes\delta{\bf B}_{1} \;-\frac{}{} (\delta\vb{\xi}_{1}\btimes{\bf B}_{0})\btimes{\bf B}_{1}\right] \\
 &  &+\; \frac{\delta{\bf B}_{1}}{4\pi} \left(\vb{\xi}_{1}\bdot{\bf B}_{0}\right) - \frac{{\bf B}_{1}}{4\pi} \left(\delta\vb{\xi}_{1}\bdot{\bf B}_{0}\right),
\end{eqnarray*}
into Eq.~\eqref{eq:id_def}, we obtain Eq.~\eqref{eq:self_adjoint}: $ \vb{\xi}_{1}\bdot\delta{\bf F}_{1} - \delta\vb{\xi}_{1}\bdot{\bf F}_{1} \equiv \nabla\bdot\delta{\bf R}_{2}$, where $\delta{\bf R}_{2}$ is given in Eq.~\eqref{eq:delta_R2}. A useful application of the identity \eqref{eq:self_adjoint} is the relation
\begin{eqnarray}
{\rm Im}\left(\vb{\xi}_{1}^{*}\bdot{\bf F}_{1}\right) & \equiv & \frac{1}{2i}\left( \vb{\xi}_{1}^{*}\bdot{\bf F}_{1} \;-\frac{}{} \vb{\xi}_{1}\bdot{\bf F}_{1}^{*}\right) \nonumber \\
 & = & \nabla\bdot\left[ \frac{1}{2i}\left( p_{1}^{*}\,\vb{\xi}_{1} \;-\frac{}{} p_{1}\;\vb{\xi}_{1}^{*}\right) \;+\; \frac{1}{8\pi i} \left[ (\vb{\xi}_{1}^{*}\btimes{\bf B}_{0})\btimes{\bf B}_{1} \;-\frac{}{} (\vb{\xi}_{1}\btimes{\bf B}_{0})\btimes{\bf B}_{1}^{*}\right] \;+\; \frac{{\bf B}_{0}}{8\pi i} \left(\vb{\xi}_{1}\btimes\vb{\xi}_{1}^{*}\right)\bdot\nabla\btimes{\bf B}_{0} \right] \nonumber \\
  & = & \nabla\bdot{\rm Im}\left[ p_{1}^{*}\,\vb{\xi}_{1} \;+\; \frac{1}{4\pi}\;(\vb{\xi}_{1}^{*}\btimes{\bf B}_{0})\btimes{\bf B}_{1} \;+\; \frac{{\bf B}_{0}}{8\pi} \left(\vb{\xi}_{1}\btimes\vb{\xi}_{1}^{*}\right)\bdot
  \nabla\btimes{\bf B}_{0} \right] \;\equiv\; \nabla\bdot\ov{\vb{\Gamma}}_{\rm MHD},
 \end{eqnarray}
 which appears in the kinetic-MHD wave-action conservation law \eqref{eq:action_kMHD}.
 \end{widetext}
 
 \acknowledgments
 
The Author wishes to acknowledge useful discussions with Dr. C. Chandre. The Author is also grateful for the several comments made by the anonymous referee during the review process. The present work was funded by a grant from the U.S. Department of Energy under contract DE-SC0014032 and was partially funded by a grant from the National Science Foundation under contract PHY-1805164.

\end{document}